\documentclass[fleqn,usenatbib]{mnras}
\usepackage{newtxtext,newtxmath}

\usepackage[T1]{fontenc}
\usepackage{ae,aecompl}

\usepackage{graphicx}	
\usepackage{amsmath}	
\usepackage{textcomp}
\usepackage{graphics}
\usepackage{longtable}[1]
\usepackage{lscape}
\usepackage{color}
\usepackage{xcolor}
\usepackage{rotating}
\usepackage{cite}
\usepackage{ulem}
\usepackage{epstopdf}
\usepackage[title,titletoc]{appendix}
\usepackage{natbib} 
\usepackage{journal_names}
\usepackage{multicol,float}
\usepackage{epstopdf}
\newcommand{\m}{$-$}
\newcommand{\p}{$+$}
\defcitealias{2013MNRAS.431..554R}{Paper I}

\title[The SUNBIRD survey: the $K$-band CLFs]{The SUNBIRD survey: the $K$-band luminosity functions of young massive clusters in intensely star-forming galaxies}

\author[Z. Randriamanakoto et al.]{ Z. Randriamanakoto$^{1,\,2}$\thanks{E-mail: zara@saao.ac.za}, P. V\"ais\"anen$^{1,\,3}$, P. Ranaivomanana$^{4}$, R. Ramphul$^{1}$, 
\newauthor
E. Kankare$^{5}$, S. Mattila$^{5}$, S. D. Ryder$^{6,\,7}$ and J. Kotilainen$^{5,\,8}$\\
$^{1}$\,South African Astronomical Observatory, P.O. Box 9, Observatory 7935, South Africa\\
$^{2}$\,Department of Physics, University of Antananarivo, P.O. Box 906, Antananarivo, Madagascar \\
$^{3}$\,Southern African Large Telescope, P.O. Box 9 Observatory, Cape Town, South Africa \\
$^{4}$\,Department
of Astrophysics/IMAPP, Radboud University, P.O.
9010, 6500 GL, Nĳmegen, The Netherlands\\
$^{5}$\,Department of Physics and Astronomy, University of Turku, FI-20014 Turku, Finland \\ 
$^{6}$\,School of Mathematical and Physical Sciences, Macquarie University, NSW 2109, Australia \\ 
$^{7}$\,Astronomy, Astrophysics and Astrophotonics Research Centre, Macquarie University, Sydney, NSW 2109, Australia \\ 
$^{8}$\,Finnish Centre for Astronomy with ESO (FINCA), University of Turku, Quantum, Vesilinnantie 5, FI-20014, Finland \\ 
}
\date{Accepted 2022 April 14. Received 2022 April 13; in original form 2021 December 10}

 \pubyear{2022}

\begin{document}
\normalem
\label{firstpage}
\pagerange{\pageref{firstpage}--\pageref{lastpage}}
\maketitle

\begin{abstract}
Strongly star-forming galaxies are prolific in producing the young and most massive star clusters (YMCs) still forming today. This work investigates the star cluster luminosity functions (CLFs, $dN/dL \propto L^{-\alpha}$) of 26 starburst and luminous infrared galaxies (LIRGs) taken from the SUNBIRD survey. The targets were imaged using near-infrared (NIR) $K$-band adaptive optics systems. Single power-law fits of the derived CLFs result in a slope $\alpha$ ranging between 1.53 and 2.41, with the median and average of $1.87 \pm 0.23$ and $1.93 \pm 0.23$, respectively. Possible biases such as blending effects and the choice of binning should only flatten the slope by no more than $\sim 0.15$, especially for cases where the luminosity distance of the host galaxy is below 100 Mpc. Results from this follow-up study strengthen the conclusion from our previous work: the CLF slopes are shallower for strongly star-forming galaxies in comparison to those with less intense star formation activity. There is also a (mild) correlation between $\alpha$ and both the host galaxy's star formation rate (SFR) { and SFR density ($\Sigma_{\rm SFR}$), i.e. the CLF flattens with an increasing SFR and $\Sigma_{\rm SFR}$.} Finally, we also find that CLFs on sub-galactic scales associated with the nuclear regions of cluster-rich targets (N $\approx$ 300) have typically shallower slopes than the ones of the outer field by $\sim 0.5$. Our analyses suggest that the extreme environments of strongly star-forming galaxies are likely to influence the cluster formation mechanisms and ultimately their physical properties.

\end{abstract}

\begin{keywords}
galaxies: interactions - galaxies: star clusters: general - infrared: galaxies.
\end{keywords}


\section{Introduction}
The study of strongly star-forming galaxies is key to understanding the evolution of star formation (SF) in galaxies. In addition to representing the most violent starbursting events and interactions in the local universe, those with infrared (IR) luminosities $L_{\rm IR}[8 - 1000\,\mu {\rm m}] = 10^{11 -12} L_{\odot}$,  known as luminous infrared galaxies (LIRGs), are
rare in the nearby universe but they dominate the star formation rate (SFR) density at $z \sim 1$ \citep[e.g.][]{2005ApJ...632..169L,2007ApJ...660...97C, 2009A&A...496...57M}. LIRGs and gas-rich starburst galaxies\footnote{There is no rigorous definition of a starburst galaxy, but in this work, any star-forming galaxy with an IR luminosity in the range $ 10.6 \lesssim {\rm log}\,(L_{\rm IR}/L_{\odot}) < 11$ and that is characterized by the ongoing extreme SF activity with a SFR of $\sim$ 5 - 50\,M$_{\odot}\,{\rm yr}^{-1}$ within the galaxy's $\sim$\,1 kpc central region falls under this category of galaxies \citep[see][]{2005ASSL..329....3H}.}, which we refer to here as strongly star-forming galaxies, are thus vital tools in the reconstruction of the cosmic star formation history \citep{ 2005ARA&A..43..727L,2014ARA&A..52..415M}. 
Furthermore, the fact that the main source of the IR luminosity is produced by strong SF bursts \citep{1996ARA&A..34..749S}, although an active galactic nucleus (AGN) may also be present \citep{2010ApJ...709..884Y}, make them ideal systems to probe 
the physical processes responsible for their intense SF activity \citep[e.g.][]{2006ApJ...650..835A,2013ApJS..206....1S,2021A&ARv..29....2P}. Under such extreme conditions, they contain large amounts of dense gas in collapsing clouds that are essential in fuelling the birth of massive star clusters (SCs), commonly known as young massive clusters \citep[YMCs, e.g.][]{1995AJ....110.2665M,1999AJ....118.1551W,2010ARA&A..48..431P,2018ASSL..424...91A,2019MNRAS.482.2530R,2020MNRAS.499.3267A}.

With their masses spanning between $\approx\,10^{4} - 10^8\, {\rm M}_{\odot}$, YMCs represent the most massive and extreme form of SF in galaxies (see e.g. \citealp{1995AJ....110.2757E}). { Tightly bound clusters with densities} $\gtrsim 10^4\,{\rm M}_{\odot}\,{\rm pc}^{-3}$ are usually young objects of ages $\lesssim  100\,{\rm Myr}$, each with radii of a few pc \citep{2010ARA&A..48..431P,2014prpl.conf..291L}. By exhibiting physical properties similar to those of globular clusters (GCs) in terms of mass and stellar density, YMCs are often believed to be GC progenitors 
\citep[e.g.][]{1992AJ....103..691H,2014prpl.conf..291L,2015MNRAS.454.1658K,2017ApJ...841..131A}. In addition, these peculiar objects can serve as natural laboratories for fine-tuning the general theory of SF mechanisms in galaxies. This is because most stars tend to be clustered in groups or associations at birth before dispersing over time (for those within loosely unbound structures) to become field population  \citep[e.g.][]{2003ARA&A..41...57L,2010ApJ...725.1717G}. Finally, initially thought to only form in the extreme environments of gas-rich starbursts and interacting LIRGs, they have also been seen in more quiescent environments, such as gas-poor normal spirals \citep{2002AJ....124.1393L,2014AJ....147...78W},
in nuclear star-forming rings of otherwise unremarkable galaxies \citep[e.g.][]{2014ApJ...797L..16V}, and nearby dwarf galaxies \citep[e.g.][]{2019MNRAS.484.4897C}. All these findings show that YMCs are good tracers of recent massive SF. Investigating their physical properties is thus essential to constrain the SF history (SFH) of their host galaxies.

Despite the importance of YMCs and the huge progress over the last decade in probing these massive clusters, their formation process is, however, not fully understood. In particular, while the clusters are discovered in a wide variety of environments, the impact of the host environment on their earliest evolutionary stages and hence their properties are still unclear. These have sparked the debate on the universality of the SC formation mechanism (e.g. \citealt{2012MNRAS.419.2606B,2014MNRAS.440L.116S,2015ApJ...810....1C,2016ApJ...826...32M,2017ApJ...849..128C,2019MNRAS.484.4897C,2019MNRAS.482.2530R,2020ApJ...888...92L}; see \citealt{2018ASSL..424...91A} for a recent review). It is not known to what extent is the role of external factors, such as the host environment, governing the life cycle of young SCs. 

To help address these issues, there has been much interest in investigating the possible links between the cluster properties and the host's global properties
\citep{2011AJ....142...79M,2014AJ....147...78W,2016MNRAS.462.3766C,2020ApJ...888...92L}. For instance, both theoretical and observational works by e.g. \citet{2012MNRAS.426.3008K}, \citet{2018MNRAS.477.1683M}, and references therein, have concluded that the cluster formation efficiency $\Gamma$ (or CFE, the fraction of stars to form in bound structures) increases with the host SFR density ($\Sigma_{\rm SFR}$, SFR normalized by the projected area of the galaxy). This observed trend is believed to be evidence of the environmentally-dependent cluster SF process. However, while some authors have backed up such arguments by studying the cluster mass and luminosity distributions, other works have instead reported that the mechanism for forming young SCs generally remains the same regardless of the host SFR level \citep[see e.g.][]{2016ApJ...826...32M,2017ApJ...849..128C,2019MNRAS.484.4897C,2019ApJ...872...93M}.

Basic diagnostic tools such as the cluster mass function (CMF, $dN/d{\rm M} \propto {\rm M}^{-\beta}$) and the cluster luminosity function (CLF, $dN/dL \propto L^{-\alpha}$) were widely used in drawing the two opposing views \citep[see e.g.][]{2010ARA&A..48..431P,2018ASSL..424...91A,2019ARA&A..57..227K}. Reasonably well-fitted by a power-law distribution with a canonical slope of $\approx $ 2 (see e.g. \citealt{1997ApJ...480..235E,1999AJ....118.1551W,2002AJ....124.1393L}, among many others), these functions are known to reflect the shape of the underlying initial mass function (IMF). 
Although LFs bin together clusters of different ages, their power-law slopes $\alpha$ are key parameters in constraining the influence of the galactic environments on the YMCs. Any deviation from $\alpha \approx 2$ should therefore be carefully investigated. This is also valid if a broken power-law or a Schechter function best represents the cluster luminosity distribution instead of the usual single power-law fit \citep{2006A&A...450..129G,2009A&A...494..539L}. To date, most of the CLF works in the literature were based either on optical observations and/or on host galaxies with luminosity distances $D_L \lesssim$ 20 Mpc to avoid dealing with resolution bias. Such choices  hinder respectively the detection of young SC candidates still deeply embedded in the dusty nuclear regions and the study of potential cluster-rich galaxies lying at larger distances.

In our previous work, we thus considered a representative sample of 8 local LIRGs imaged with near-IR adaptive optics (NIR AO) instruments to investigate the properties of massive star clusters in the galactic environments of strongly star-forming galaxies (\citealt{2013MNRAS.431..554R},  hereafter referred to as \citetalias{2013MNRAS.431..554R}). This was achieved by deriving the $K_s$-band\footnote{Hereafter, we will refer to both the Johnson $K$-band and the 2MASS $K_s$-band as $K$-band.} CLFs of the targets that are part of the SUNBIRD survey (SUperNovae and starBursts in the InfraReD or Supernovae UNmasked By InfraRed Detection, \citealp{2014mysc.conf..185V,2018MNRAS.473.5641K}). 
The NIR AO observing strategy was adopted to minimize effects from dust extinction while correcting for atmospheric turbulence \citep[see e.g.][]{2007ApJ...659L...9M}.
{ We found that galaxies with extreme SF activity such as interacting LIRGs are associated with  shallower power-law slopes ($\alpha \approx 1.9$)
compared to those of low SFR galaxies.} A similar trend was observed by other CLF studies of LIRGs \citep[e.g.][]{2016MNRAS.462.3766C,2020ApJ...888...92L}, which interpreted the flattening in the high-end CLF slope as a possible imprint of the host's extreme environment on the YMC properties. In contrast, SC analyses by e.g. \citet{2011PhDT.........8V}, \citet{2014AJ....147...78W} and \citet{2016ApJ...826...32M} have suggested that the discrepancy in the value of $\alpha$ largely results from resolution bias and simple statistics. However, our comprehensive blending analysis 
showed that such factors only decrease the value of $\alpha$ by $\sim$0.1 for targets at distances of $D_L <$ 100 Mpc when appropriately sized photometric apertures are used. 

The YMC study by \citetalias{2013MNRAS.431..554R} is unique in using a representative sample of galaxy hosts with high SFRs (SFR $> 30\,{\rm M}_{\odot}\,{\rm yr^{-1}}$, with median $D_L \sim$ 70 Mpc).  Discussion of the findings was however limited partially because observations were made in a single filter, and especially due to the small sample size preventing robust correlation searches with the global properties of the host galaxies.
Considering a larger sample whose cluster luminosities have been uniformly derived is thus highly advantageous to provide improvements over our pilot study and to also help address, at least to a first order, the differing views on the link between YMCs and their host galaxy environments.

The current paper is thus a follow-up study to our work published in \citetalias{2013MNRAS.431..554R}. To further assess 
findings from our pilot study, particularly the impact of the host galaxy environments on the characteristics of their YMCs, { we compile the $K$-band CLFs of 26 SUNBIRD targets, in addition to the 8 galaxies from the original sample of \citetalias{2013MNRAS.431..554R}}. This much larger sample of strongly-star forming galaxies was observed in the NIR regime using the Very Large Telescope NAOS-CONICA \citep[VLT/NACO,][]{2003SPIE.4841..944L,2003SPIE.4839..140R} AO systems.
This work also aims to investigate whether the global properties of the overlapping sample studied by \citet{2018PhDT.......244R} are somehow physically related with the derived slopes of the individual and composite-based NIR CLFs. Our ultimate goal is therefore to 
constrain the formation process of these massive stellar clusters that are residing in the realm of starburst-dominated galaxies.

The paper is structured as follows. Section\,\ref{sec:sunbird} describes the SUNBIRD survey and the sample we used in this work. We briefly present the observations, the source photometry and the star cluster catalogues in Section\,\ref{sec:catalogue}. The results, including the derived NIR CLFs and the correlation search between $\alpha$ and the galaxy global properties, are reported respectively in Sections\,\ref{sec:CLFs} and \ref{sec:search}. We discuss our findings in Section\,\ref{sec:discuss} and then draw our conclusions in Section\,\ref{sec:conclusion}.

\section{The SUNBIRD survey}\label{sec:sunbird}

\subsection{Survey description}
Our parent sample is { a set of 42} strongly star-forming galaxies from the SUNBIRD survey. This ongoing project uses ground-based NIR telescopes equipped with AO imaging to observe a representative sample of starburst galaxies and LIRGs in the nearby universe \citep{2014mysc.conf..185V,2018MNRAS.473.5641K}. SUNBIRD is mainly designed to use the NIR AO capabilities to search for optically hidden core collapse SNe (CCSNe) that are expected to reside in the nuclear regions of the high SFR galaxies \citep{2001MNRAS.324..325M}. High spatial resolution instruments mounted on the Gemini-North (ALTAIR/NIRI), VLT (NACO),  Gemini-South (GeMS/GSAOI), and Keck II (NIRC2) telescopes were used to image the SUNBIRD targets. Coupled with low levels of dust extinction in the NIR regime (reduced by a factor of 10 compared to that of the optical range), such observations are efficient for the detection of SNe missed by traditional optical   surveys \citep[see e.g.][]{2007ApJ...659L...9M,2012ApJ...756..111M,2008ApJ...689L..97K,2012ApJ...744L..19K,2021A&A...649A.134K,2018MNRAS.473.5641K}. CCSNe are direct tracers of the current rate of massive star formation.
Hence, they are useful in characterizing the SFH of their host galaxies and the Universe \citep[e.g.][]{2012ApJ...757...70D}. 

Besides the detection of dust obscured CCSNe, the SUNBIRD sample is also used to study the physical details of SF activity and its triggering in the context of strongly star-forming galaxies \citep[e.g.][]{2008MNRAS.384..886V,2014mysc.conf..185V}. While  \citet{2015salt.confE..18R} and \citet{2018PhDT.......244R} conducted follow-up spectroscopic observations with the Robert Stobie Spectrograph (RSS, \citealp{Burgh2003}) on the Southern African Large Telescope (SALT, \citealp{Buckley2006}) to determine the stellar population properties of the galaxy sample (see Section\,\ref{sec:Ramphul-work}), \citet{2013ApJ...775L..38R}, \citetalias{2013MNRAS.431..554R} and \citet{2015PhDT.......214R} focused instead on photometric investigations of their YMC populations. 

The SUNBIRD parent sample is drawn from the flux-limited IRAS Revised Bright Galaxy Sample (RBGS, \citealp{2003AJ....126.1607S}). The targets were selected based on the following criteria: {\it i)} they have an IR luminosity range of 
$10.6 \lesssim {\rm log\,(L_{IR}/L_{\odot})} \lesssim 11.9$; {\it ii)} the redshift coverage is up to $z \sim 0.05$ which translates to a luminosity distance $D_L \lesssim 200$\,Mpc; {\it iii)} even though AGN were not excluded a priori, the sample is predominantly composed of starburst-dominated systems with cool 
IRAS colours where $f_{25}/f_{60} < 0.2$; {\it iv)} there is a suitable bright star nearby to serve as the AO natural guide star (NGS). 
These requirements resulted in a representative  statistical sample of IR-bright galaxies within our distance limit and above a luminosity ${\rm log\,(L_{IR}/L_{\odot})} = 10.6$, with a wide range of morphologies and interaction stages (Section\,\ref{sec:morpho}).

\subsection{SUNBIRD targets studied in this work}\label{sec:sample}

\begin{figure*}
\begin{center}
\resizebox{1.0\hsize}{!}{
\includegraphics[width=\textwidth, trim= 0cm 0cm 0cm 0cm]{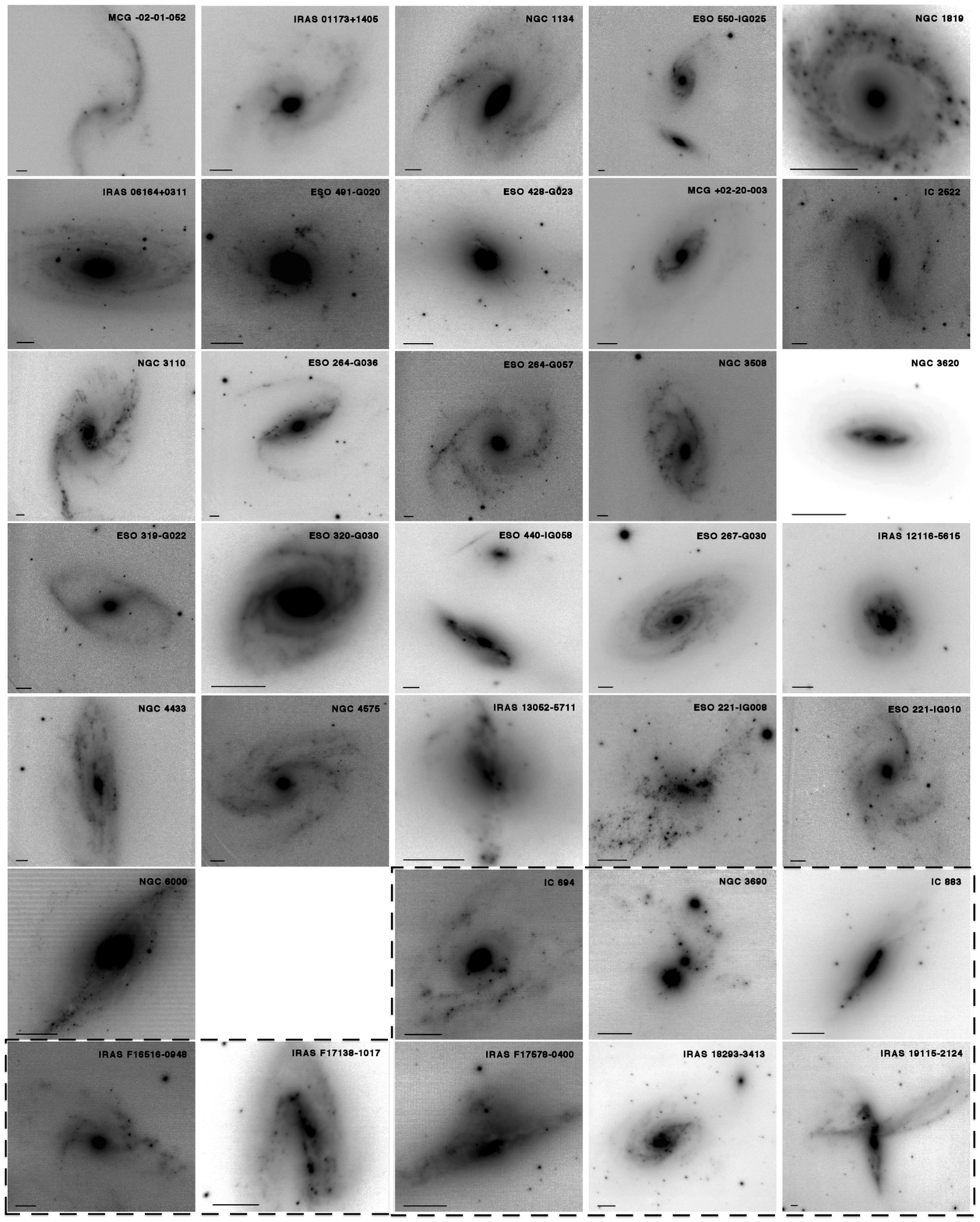}
}
\end{center}
\caption{\small NIR AO images for the { 34 SUNBIRD targets studied in this work, 8 of which (demarcated by the black dashed lines) had their CLFs already published in our \citetalias{2013MNRAS.431..554R} pilot study}. The name of each galaxy is shown in each panel. The horizontal line represents a scale of 1\,kpc where the pixel scale values are respectively 0.022, 0.027, and 0.054 arcsec pixel$^{-1}$ for Gemini/ALTAIR/NIRI, VLT/NACO S27 and S54 images. North is up and East is left.} 
\label{fig:sample}
\end{figure*} 

\begin{table*}

\caption{\small The SUNBIRD galaxies mainly studied in this work (top) and those that  already served as targets for our $K$-band CLF pilot study (bottom).} 
\centering
\setlength{\tabcolsep}{10pt}
\scalebox{1.}{
\begin{tabular}{clcccc@{\hspace{7pt}}c}
\hline \hline
   \noalign{\smallskip}
IRAS name	&\hspace{0.5cm} Galaxy &	RA	&DEC  &	log\,$L_{\rm{IR}}$	&	$D_{L}$ & Interaction \\
 			&\hspace{0.7cm}name	&(J2000)	&(J2000)	& ($L_{\odot}$) 		&	(Mpc)	  &   phase        	  		\\
(1)	&	\hspace{0.8cm}(2)	&	(3)	&	(4)	&	(5)	&	(6)	&	      (7)			\\
  \noalign{\smallskip}
\hline 
   \noalign{\smallskip}
\multicolumn{7}{c}{Targets with unpublished NIR CLFs} \\
\hline 
\noalign{\smallskip}
F$00163-1069$&MCG\,\m 02-01-052	&00h18m50.1s& \m10d21m42s&10.63$^{\dagger}$  &  110.0 & II \\

~~$01173+1405$& IRAS $01173+1405$ &01h20m02.7s& \p14d21m43s &  11.63&127.0& II\\

F$02509+1248$&NGC\,1134		&02h53m41.3s &\p13d00m51s   &10.83&47.4& II\\

F$04191-1855$&ESO 550\m IG025	&04h21m20.0s& \m18d48m48s& 11.45 &  135.0&  II \\

F$05091+0508$&       NGC\,1819	&05h11m46.1s& \p05d12m02s	&10.90&	61.9	& 0\\

~~$06164+0311$& IRAS\, $06164+0311$				&06h19m02.6s& \p03d09m51s&10.79&41.5&0\\

~~$07077-2729$&ESO\,491\m G020 &07h09m48.1s &\m27d34m15s &10.86$^{\dagger}$&43.5& II \\

~~$07202-2908$&ESO\,428\m G023&07h22m09.4s& \m29d14m08s&10.76&44.5&0 \\

F$07329+1149$& MCG\,\p 02-20-003 &07h35m43.4s& \p11d42m34s	&11.08&70.5& II \\

F$09529-3253$ & IC\,2522 		&09h55m08.9s& \m33d08m14s&10.63&46.1& I\\

F$10015-0614$&NGC\,3110		&10h04m02.1s &\m06d28m29s& 11.31&75.2& II\\

F$10409-4556$&ESO\,264\m G036    &10h43m07.7s &\m46d12m45s&11.35&92.0& I \\

F$10567-4310$&ESO\,264\m G057    &10h59m01.8s  &\m43d26m26s&11.08&75.8& II \\

F$11005-1601$&NGC\,3508		&11h02m59.7s &\m16d17m22s&10.90&59.1&0\\

F$11143-7556$ & NGC\,3620	         &11h16m04.7s  &\m76d12m59s&10.70&24.9&0\\

F$11255-4120$&ESO\,319\m G022&11h27m54.1s &\m41d36m52s&11.04&72.3&0\\

F$11506-3851$&ESO\,320\m G030	&11h53m11.7s &\m39d07m49s&11.10&49.0&0\\

F$12043-3140$&ESO\,440\m IG058    &12h06m51.9s &\m31d56m54s &11.36 &102.0 & II \\

F$12115-4656$&ESO\,267\m G030     &12h14m12.9s &\m47d13m42s   &11.19&80.9& II\\

~~$12116-5615$& IRAS\, $12116-5615$			&12h14m22.1s &\m56d32m33s&11.59&117.0 & IV\\

F$12250-0800$&NGC\,4433		&12h27m38.6s &\m08d16m42s&10.87&46.3& II \\

F$12351-4015$&NGC\,4575 		&12h37m51.1s &\m40d32m14s  &10.96&45.0 & I \\

~~$13052-5711$	& IRAS\, $13052-5711$			&13h08m18.7s &\m57d27m30s&11.34&91.6 & IV\\

F$13473-4801$&   ESO 221\m IG008  &13h50m26.4s &\m48d16m36s&	10.77 & 46.7 & II\\

F$13478-4848$ & ESO\,221\m IG010	&13h50m56.9s &\m49d03m20s&11.17&45.9& II\\

F$15467-2914$&NGC\,6000		&15h49m49.5s&\m29d23m13s&10.97&32.1& IV   \\

  \noalign{\smallskip}
\hline 
   \noalign{\smallskip}
\multicolumn{7}{c}{Targets with published NIR CLFs} \\
\hline 
\noalign{\smallskip}

F$11257+5850$&NGC\,3690E$^{\dag}$		&11h28m27.3s &\p58d34m43s&11.66$^{\ddagger}$&45.3 & III \\

F$11257+5850$& NGC\,3690W$^{\dag}$		&11h28m32.3s&\p58d33m43&11.48$^{\ddagger}$&45.3 & III \\

F$13182+3424$& IC 883$^{\dag}$		        &13h20m35.3s &\p34d08m22s&11.67&101.0& IV\\

F$16516-0948$&IRAS\, F$16516-0948$$^{\dag}$			&16h54m24.0s& \m09d53m21s&11.24&94.8 & IV\\

F$17138-1017$&IRAS\, F$17138-1017$$^{\dag}$			&17h16m35.8s&\m10d20m39s&11.42&72.2& III \\

F$17578-0400$&IRAS\, F$17578-0400$$^{\dag}$			&18h00m31.9s& \m04d00m53s&11.35&57.3& II \\

~~$18293-3413$&IRAS\, $18293-3413$			&18h32m41.1s& \m34d11m27s&11.81&74.6& II \\

~~$19115-2124$& IRAS\, $19115-2124$	&19h14m30.9s&\m21d19m07s&11.87&206.0 & III \\

\noalign{\smallskip}
\hline
\noalign{\smallskip}
\multicolumn{7}{@{} p{14.5cm} @{}}{\footnotesize{\textbf{Notes. } The targets are ordered with increasing RA. Col 1: {\tt IRAS} survey name; Col 2:  galaxy name, { any target marked by $\dagger$ has been imaged with Gemini/ALTAIR/NIRI, whereas the rest of the sample with VLT/NACO. In the literature, IRAS\, $19115-2124$ is also dubbed the Bird}; Cols 3 \& 4: equatorial coordinates; Col 5: galaxy IR luminosity from \citet{2003AJ....126.1607S}, any value marked by $\ddagger$ is estimated by using the method described in \citet{2013ApJ...775L..38R}; Col 6: luminosity distance retrieved from NED database; Col 7: the galaxy interaction phase where I refers to the first approach, II to the pre-merger phase, III and IV for merger and post-merger stages, respectively. Class 0 regroups galaxies with undisturbed morphologies that are apparently isolated.}}
\end{tabular}
}
\label{tab:sample}
\end{table*}

\begin{figure}
\centering
\resizebox{1.\hsize}{!}{\includegraphics{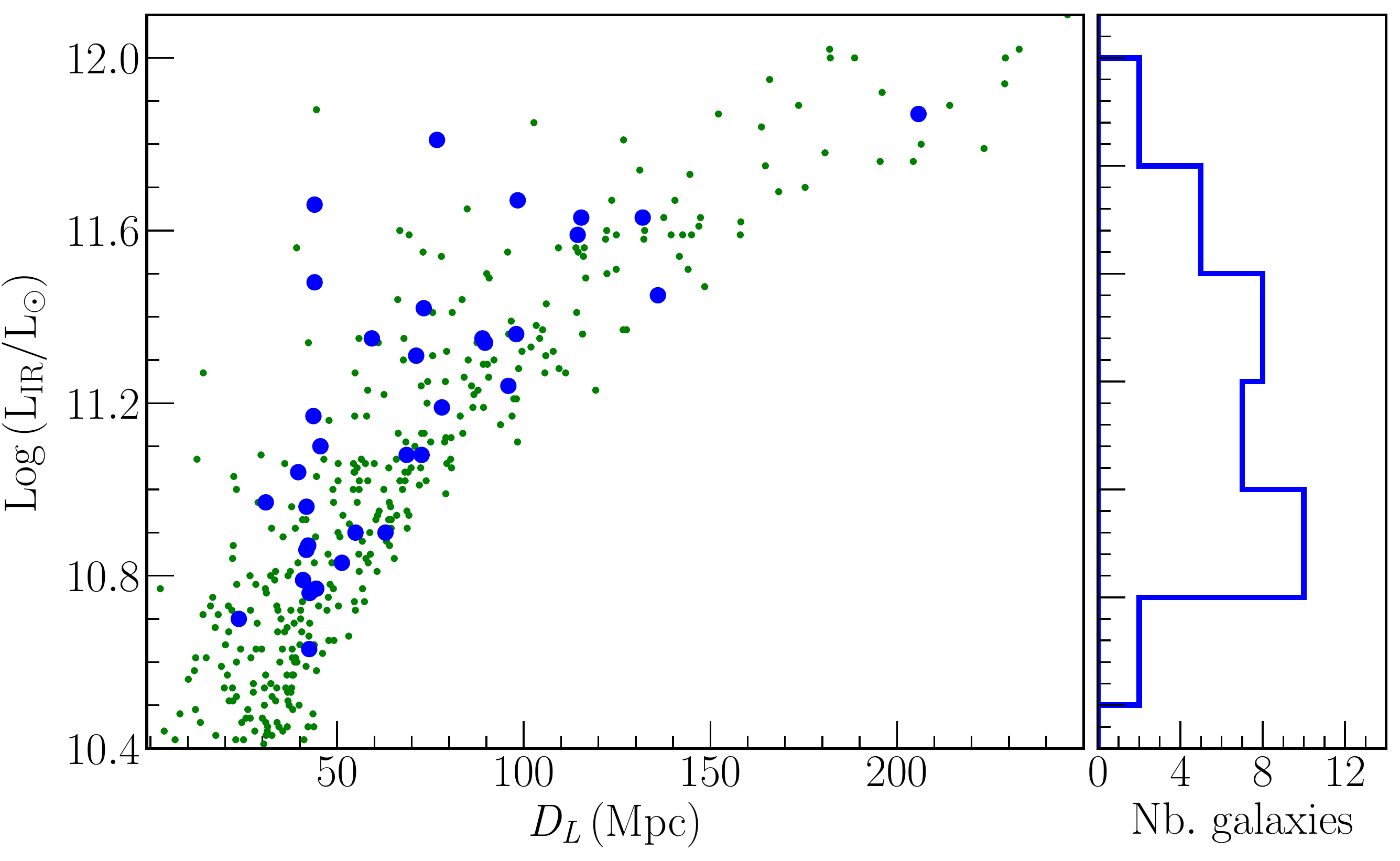}}
\caption{\small Distribution of IRAS galaxies from RBGS (\citealp{2003AJ....126.1607S};  green dots) in
the IR luminosity - distance plane. Our targets from the SUNBIRD sample are shown as blue circles and are spread approximately
homogeneously all over the 2D-plane up to  { $D_L \sim 150$} Mpc. The right panel shows a histogram of IR luminosities of the SUNBIRD galaxy sample with a bin size of 0.2 in the logarithmic scale.}
\label{fig:LirDistr}
\end{figure} 

\begin{figure}
\centering
\resizebox{.85\hsize}{!}{\includegraphics[trim= 3cm 0cm 3cm 0cm]{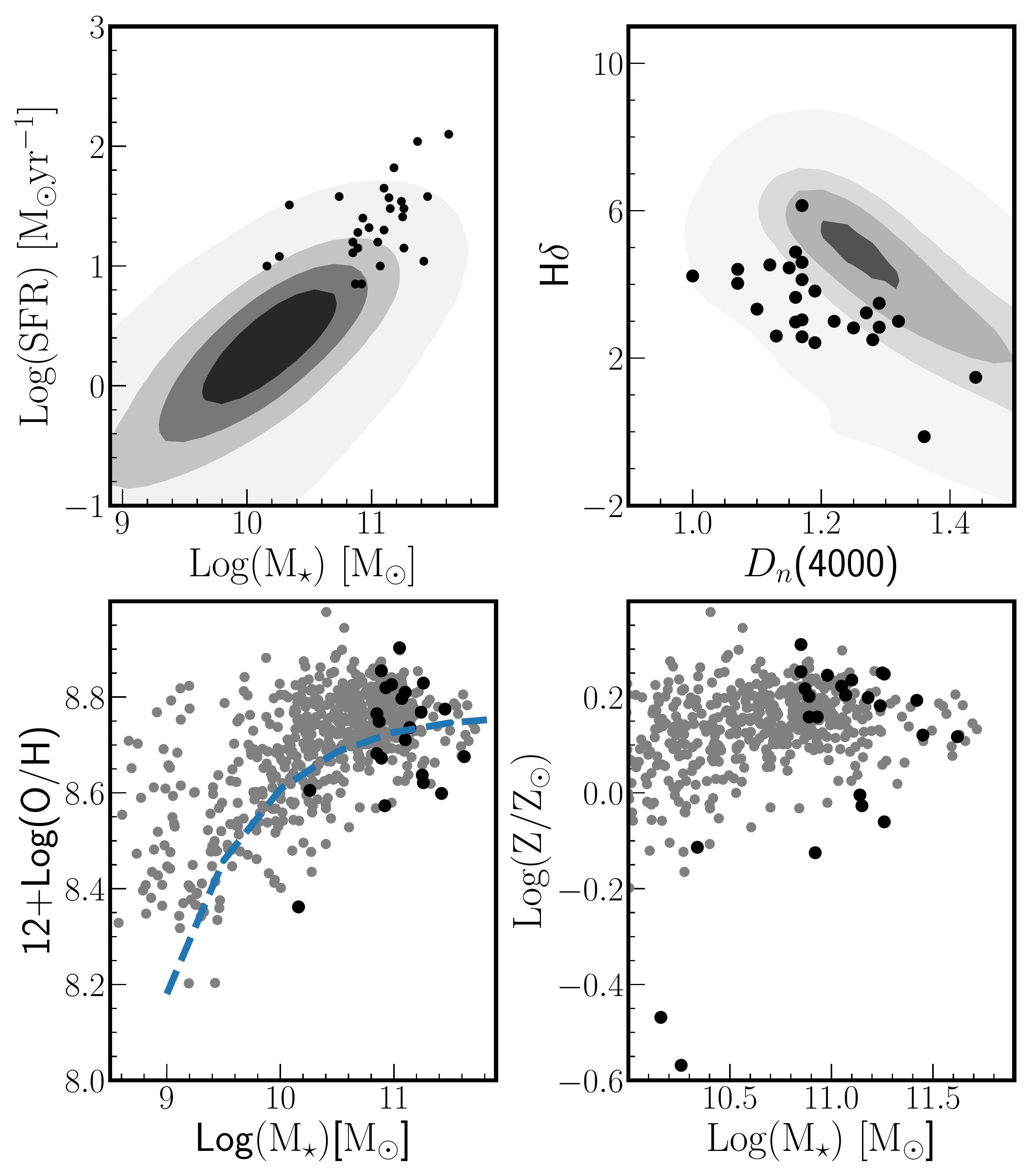}}
\caption{\small Some physical properties of the SUNBIRD sample { studied in this work }(with available data from \citealt{2018PhDT.......244R}) represented by the black points as compared to survey data { (density contour plots and grey data points for top and bottom panels respectively)} and/or models (dashed line). {\em Top panels}: the global SFR - stellar mass relation ({\em left}) and H$\delta$ plotted against $D_n(4000)$ ({\em right}) for the SUNBIRD galaxies as compared to low redshift SDSS galaxies \citep{2004MNRAS.351.1151B, 2009ApJS..182..543A}. { The dark region of the contour plot indicates the highest concentration of the local SDSS galaxies}; {\em Bottom panels}: stellar mass vs. oxygen abundance ({\em left}) and stellar mass-metallicity ({\em right}) plots comparing the SUNBIRD subsample to local CALIFA galaxies studied by \citet{2013A&A...554A..58S}. The dashed line represents the best-fitted relation considering the CALIFA sample.}

\label{fig:prop-sample}
\end{figure}

From the parent sample of \citet{2013ApJ...775L..38R}, we exclude 8 galaxies with a relatively small number of YMCs (i.e. N $ < 20$) to avoid statistical bias in the fitted CLFs. We thus end up with 34 targets, 8 of which had their CLFs already published in our \citetalias{2013MNRAS.431..554R} pilot study, whereas the rest of the galaxy CLFs are presented in this work.  By considering the derived sample, we hope to conduct a more robust assessment of any possible biases as well as our correlation searches with the CLF slope. Fig. \ref{fig:sample} displays the NIR AO images of these targets. They have been observed with VLT/NACO, except NGC 3690, IC 883, IRAS F16516$-$0948, IRAS F17138$-$1017, and IRAS F17578$-$0400 that were imaged with Gemini/ALTAIR/NIRI. 
 
The constructed sample consists of 12 gas-rich starburst galaxies and 22 LIRGs. We refer to Table\,\ref{tab:sample} for the properties of these 34 selected galaxies. To show the different IR luminosity and distance ranges covered in this work, we plot in the IR luminosity - distance plane a distribution of our SUNBIRD galaxies (blue points) on top of the parent RBGS sample labelled as green dots (see Fig.\,\ref{fig:LirDistr}). Our sample is reasonably spread homogeneously in the luminosity baseline with a median value of $1.2\,\times\,10^{11}\,L_{\odot}$. The targets have distances below 135 Mpc, except for {  IRAS\,19115$-$2124 (also known as the Bird, \citealp{2008MNRAS.384..886V}) which is located at $D_L$ = 206\,Mpc with an angular scale of 0.97 kpc arcsec$^{-1}$. Because of its relatively large distance compared to the rest of the sample, we exclude data from that target from some tests to avoid resolution bias and treat it separately when investigating statistical bias in Section \ref{sec:stat-bias}}. We note that our choice to work on the same sample as in \citet{2013ApJ...775L..38R}  is also motivated by the availability of the galaxy $K$-band star cluster catalogues ready for analysis (see Section\,\ref{sec:catalog}).

\subsubsection{Morphologies and interaction stages}\label{sec:morpho}

The sample covers a wide variety of morphologies and interaction stages: the first approach, pre-merger,
merger, and post-merger stages are respectively annotated as I, II, III and IV in Table\,\ref{tab:sample}. 
Note that the classification is based on studying the apparent morphologies of the targets observed in 
the high-spatial resolution NIR AO images 
in Fig.\,\ref{fig:sample}.

The galaxy disks remain stable during the first approach, though the gas content becomes perturbed due to the violent dynamical evolution (Class\,I).
Features such as tidal tails and bridges are indicative of pre-merger stages with the distances between the two disks and the two nuclei of the galaxies
still far enough to be detected individually (Class\,II). However, merging stages are underway when the two coalesced nuclei are separated by a
relatively small projected distance (not more than 2\,kpc) and both disks are completely distorted to allow the formation of a common internal structure (Class\,III). 
When the coalesced nuclei have merged completely, the more relaxed post-merger system has a much brighter nucleus enveloped with some shell
structures (Class IV). This customized classification scheme is a simplified version of that from \citet{2002ApJS..143..315V} and similar
to the method adopted by \citet{2011AJ....142...79M}. Finally, targets that are apparently undisturbed with no obvious pair within 10\,arcmin
radius, corresponding to 100 to 300\,kpc radius in the distance range of the bulk of the sample, are classified as isolated galaxies. They are identified
as Class\,0 in Table\,\ref{tab:sample}. In the following sample, around a quarter appear to be isolated and relaxed spirals { (Class\,0)}, though sometimes slightly perturbed { as the galaxies enter on their initial approach (Class\,I),
another quarter are currently interacting or in a post-merger stage (Class\,III and IV), and the rest are morphologically disturbed objects during pre-merger stage (Class\,II)}.   

\subsubsection{Global galaxy properties}\label{sec:Ramphul-work}

We highlight in this section relevant global properties of our galaxy sample. The properties listed in Table\,\ref{tab:prop-Ramphul} are fully presented in \citet{2018PhDT.......244R} and V\"ais\"anen et al. (in prep); we also refer to \citet{2015salt.confE..18R} and \citet{2017geat.confE..29R}.  Very briefly, full-spectrum stellar population analysis was done on low and medium resolution long-slit spectra obtained on the SALT/RSS\footnote{Data were gathered from June 2011 until June 2014 under the following programs: 2011-3-RSA\_OTH-023, 2012-1-RSA\_OTH-032, 2012-2-RSA\_OTH-015, 2013-1-RSA\_OTH-024, 2013-2-RSA\_OTH-006, 2014-1-RSA\_OTH-002.} using STARLIGHT \citep{2005MNRAS.358..363C} fitting procedures based on the \citet{2003MNRAS.344.1000B} BC03 library of synthetic Single Stellar Population (SSPs). Spatially resolved (i.e.\ along the slit) stellar population characteristics were derived, including ages, Oxygen abundances and metallicities, extinction properties, as well as both stellar and gas kinematics, and more detailed SFHs.  

A selection of integrated galaxy properties is considered here for the purpose of searching for correlations with the YMC characteristics. { In this work, we use global SFR values} estimated based on the galaxy IR luminosity \citep{1998ARA&A..36..189K}. { This SFR indicator is associated with regions of age below 100 Myr where young and extreme SF bursts are responsible for the large fraction of IR emission from starburst galaxies and LIRGs \citep[e.g.][]{2012ARAA..50..531K,2021A&ARv..29....2P}. The derived SFR levels thus cover the current/recent SFH of the host galaxy. They probe similar timescales to those reconstructed from YMC studies, since these massive objects form whenever there is intense SF activity, which makes them a good tracer of small-scale SF mechanisms.}

The stellar mass M$_\star$ (derived from 2MASS $K$-band luminosity while taking into account Galactic extinction effects and doing {\it k}-correction), 
and the specific SFR, sSFR = SFR/M$_\star$ are also used, along with the STARLIGHT-derived light ({\it l}) and mass ({\it m}) weighted best-fit stellar population ages and stellar metallicities of the galaxies. These parameters are denoted as $({\rm Age})_l$, $({\rm Age})_m$, $(Z)_l$, $(Z)_m$, respectively. We also use  H$\delta$ and the D4000 index (an indicator of the 4000\AA\ break) as measured directly from the spectra. Numerous emission line strengths were measured after the best fit stellar continuum was subtracted from the observed spectra. Table\,\ref{tab:prop-Ramphul} shows the measured Equivalent Width (EW) of H$\alpha$, and the Oxygen abundance measured from a variety of strong emission line diagnostics calibrated to a common O3N2 base following the methods of \citet{2008ApJ...681.1183K}. { Parameters such as $D_n(4000)$ and H$\delta$ can be used as a proxy for the age of the stellar population while EW(H$\alpha$) is a good indicator of recent SF.}

Fig. \ref{fig:prop-sample} shows some relevant integrated properties of the SUNBIRD { galaxies that are studied in this work} with respect to their stellar mass, SFR and Oxygen abundance characteristics. { These parameters are taken from \citet{2018PhDT.......244R}}. The top panels overplot the SFR - stellar mass relation and H$\delta$ vs. $D_n(4000)$ of the SUNBIRD subsample and local star-forming SDSS galaxies from the literature \citep{2004MNRAS.351.1151B, 2009ApJS..182..543A}. The bottom panels compare the stellar mass vs. Oxgen abundance and stellar mass-metallicity of the SUNBIRD galaxies to the properties of local CALIFA galaxies studied by \citet{2013A&A...554A..58S}. For all four plots, our targets generally fall on top of the distribution of low redshift SDSS and CALIFA galaxies,  placing the SUNBIRD galaxies in the context of the galaxy population as a whole. Overall, they are massive galaxies with high SFR levels and with an underabundance in metallicity by $\sim$ 0.1 dex. The two outliers in the stellar mass-metallicity relation (bottom right) correspond to ESO 221\m IG008 and ESO 491\m G020, which have their stellar masses relatively lower than the median value of log\,(M$_{\star}/{\rm M}_{\odot}) \sim$ 11.08 for the SUNBIRD sample. The pre-merging process happening in these targets is likely to induce inflows of metal-poor gas, and once  mixed with the enriched gas would lower the observed metallicity whilst triggering new SF episodes.  Finally, most of the SUNBIRD targets are associated with relatively low values of H$\delta$ and $D_n(4000)$ compared to SDSS galaxies, i.e. they are located within $\sim$\,3$\sigma$ of the distribution shown in the top right panel. This is due to the very young SF episodes characterising the SUNBIRD galaxies. Extensive physical interpretations of the correlations shown in Fig. \ref{fig:prop-sample} can be found in \citet{2018PhDT.......244R}.

\section{NIR Data and source catalogues}\label{sec:catalogue}

\subsection{Observations and photometry}
We used the VLT NACO instrument to obtain the $K$-band AO images of the 26 main targets between October 2010 and June 2011 (PI: Escala, PID:\,086.B-0901).
Some of the targets overlapped with a sample from the observing run of PID:\,089.D-0847 (PI: Mattila). Therefore, complementary data
from the latter cycle were also included in this work. Either S27 or S54 cameras was used, taking into account
the size of the galaxy, providing a plate scale of 0.027 or 0.054\,arcsec pixel$^{-1}$, respectively.
Individual frames were taken in dithering mode with 120\,s per pointing. With a point spread function (PSF) with FWHM of $\sim$\,0.1\,arcsec { (equivalent to a physical size of $\sim 12 - 60$ pc)}, the final science images have resolutions that match with observations from the HST. The total on-source integration times range between 20 and 40 minutes.  We refer to  \citet{2013ApJ...775L..38R} and \citetalias{2013MNRAS.431..554R} for more details on our {\tt IRAF}-based data reduction pipeline. In the case of NGC\,6000 and NGC\,6240, NIR
images were acquired from the observing runs of PIDs:\,084.D-0261 and 087.D-0444 (PI: Mattila). { Given that YMCs have typical sizes of $3 - 5$ pc \citep[e.g.][]{1999AJ....118.1551W,2021MNRAS.508.5935B}, we expect that our sample would contain both individual clusters and small stellar complexes, i.e. clusters of clusters on scales $\sim 20 - 50$ pc.}

Object detection using {\tt SEXtractor} \citep{1996A&AS..117..393B} was performed on the unsharped-masked
version of the images. { A minimum of $8 - 10$ contiguous pixels above threshold combined with a detection limit of $\sim1.5\sigma$ above the rms background were chosen to detect potential candidates.} We then applied aperture photometry on the catalogue with aperture
radii of 2 and 3 pixels (0.11 and 0.08\,arcsec) for S27 and S54 frames, respectively. 
Sky annuli were $0.05 - 0.08$\,arcsec in width { with an inner radius of one pixel away from the aperture radius in both cases}. Depending on the number of isolated point sources in the field,
we either derived a constant (growth curves until 1 arcsec) or an AO-distance dependent aperture correction $a_c$ to account for the small aperture sizes  of 2 and 3 pixels which respectively recover around 17 and 39 percent of the source total flux, i.e. $a_c \sim 1.03 - 1.95$ mag. If recorded, {\tt VEGAMAG} zero-points $m_0$ were taken from the ESO/NACO official website. Otherwise, the same procedure as in \citetalias{2013MNRAS.431..554R} was adopted to estimate $m_0$. 
The uncertainty of the absolute magnitudes ranges between $\approx 0.1 - 0.3$\,mag.

\begin{table*} 

\caption{\small Global properties of the targets derived from stellar population analysis.} 
\label{tab:prop-Ramphul}
\centering
\resizebox{1.\textwidth}{!}{%
\renewcommand{\arraystretch}{1.}
\begin{tabular}{lcccccccccccc}   
\hline \hline
\noalign{\smallskip}  
\hspace{0.5cm} Galaxy name  & log\,${\rm M_\star}$ & log\,${\rm (SFR/M_\star)}$ & EW(H$\alpha$)&$A_{\rm v, st}$ & $A_{\rm v, HII}$ &D4000&H$\delta$&{ 12+$\log$(OH)}& log\,(Age)$_l$ & log\,(Age)$_m$ & $Z_l$ & $Z_m$  \\ 
&(${\rm M}_{\odot}$)&$({\rm yr}^{-1})$&&(mag)&(mag)&&&&(yr)&(yr)&($Z_{\odot}$)&($Z_{\odot}$)\\
\hspace{1.1cm} (1)	& (2)	&	(3)	&	(4)	&	(5)	&	(6)	&	      (7) & (8) & (9) & (10) & (11) & (12) & (13)	\\
\noalign{\smallskip}       
\hline \hline
\noalign{\smallskip} 
MCG\,\m 02$-$01$-$052 & 10.92 & $-$10.07 & 74.56 & 0.45 $\pm$ 0.02 & 0.87 $\pm$ 0.02 &
1.07 & 4.03&8.57 & 7.68 $\pm$ 0.02 & 9.86 $\pm$ 0.06 & 0.88  $\pm$  0.07 & 0.75  $\pm$ 
0.20\tabularnewline
IRAS\, 01173\p1405 & - & - & - & - & - & - &
- & - & - & - & - & -\tabularnewline
NGC 1134 & - & - & - & - & - & - & - & -
& - & - & - & - \tabularnewline
ESO\, 550\m IG025  & 11.15 & $-$9.67 & - & 1.48  $\pm$  0.04 & 2.19  $\pm$ 
1.41 & 1.07 & 4.41 &-& 7.34  $\pm$  0.03 & 9.83  $\pm$  0.06 & 1.06  $\pm$  0.08 & 0.94
 $\pm$  0.07\tabularnewline
NGC 1819  & 11.26 & $-$10.11 & 19.71 & 0.58 $\pm$ 0.02 & 1.77 $\pm$ 0.03 & 1.27 &
3.23 &8.83& 8.52 $\pm$ 0.02 & 9.91 $\pm$ 0.02 & 0.81  $\pm$  0.05 & 1.77  $\pm$ 
0.08\tabularnewline
IRAS\, 06164\p0311 & 11.42 & $-$10.38 & 2.86 & 0.74 $\pm$ 0.07 & 1.67 $\pm$ 0.87
& 1.36 & $-$0.13 & 8.60& 8.96 $\pm$ 0.07 & 10.10 $\pm$ 0.01 & 0.99  $\pm$  0.13 & 1.56  $\pm$ 
0.10\tabularnewline
ESO\, 491\m G020 & 10.26 & $-$9.18 & 36.11 & 0.17 $\pm$ 0.04 & 1.29 $\pm$ 0.04 &
1.17 & 2.58 &8.61& 8.40 $\pm$ 0.01 & 9.82 $\pm$ 0.07 & 0.40  $\pm$  0.08 & 0.27  $\pm$ 
0.09\tabularnewline
ESO\, 428\m G023 & 11.07 & $-$10.07& 21.45 & 0.61 $\pm$ 0.03 & 2.08 $\pm$ 0.03 &
1.29 & 2.84&8.80 & 8.62 $\pm$ 0.03 & 9.88 $\pm$ 0.03 & 0.64  $\pm$  0.06 & 1.60  $\pm$ 
0.13\tabularnewline
MCG\,\p 02$-$20$-$003 & - & - & - & - & - & - & - & - & - & - & - &  - \tabularnewline
IC\, 2522 & 10.87 & $-$10.02 & 21.33 & 0.80 $\pm$ 0.03 & 1.76 $\pm$ 0.03 & 1.19 &
3.82 & 8.75&8.06 $\pm$ 0.05 & 9.94 $\pm$ 0.04 & 0.59  $\pm$  0.07 & 1.65  $\pm$ 
0.18\tabularnewline
NGC 3110 & 11.24 & $-$9.07 & 41.47 & 1.19 $\pm$ 0.02 & 2.28 $\pm$ 0.02 & 1.17 &
4.13 & 8.77&7.87 $\pm$ 0.02 & 9.89 $\pm$ 0.03 & 0.58  $\pm$  0.04 & 1.52  $\pm$ 
0.13\tabularnewline
ESO\, 264\m G036 & 11.45 &$-$9.87 & 15.46 & 0.67 $\pm$ 0.02 & 3.07 $\pm$ 0.05 &
1.29 & 3.49&8.77 & 8.46 $\pm$ 0.04 & 9.81 $\pm$ 0.04 & 0.72  $\pm$  0.05 & 1.32  $\pm$ 
0.12\tabularnewline
ESO\, 264\m G057 & 11.10 & $-$9.08 & 37.32 & 1.29 $\pm$ 0.03 & 2.81 $\pm$ 0.09 &
1.17 & 3.04 &8.81& 7.77 $\pm$ 0.04 & 9.97 $\pm$ 0.03 & 0.75  $\pm$  0.06 & 1.72  $\pm$ 
0.11\tabularnewline
NGC 3508 & 10.89 & $-$9.74 & 36.20 & 1.12 $\pm$ 0.03 & 1.91 $\pm$ 0.02 & 1.15 &
4.45 &8.67& 7.98 $\pm$ 0.03 & 9.91 $\pm$ 0.03 & 0.68  $\pm$  0.05 & 1.59  $\pm$ 
0.13\tabularnewline
NGC 3620  & - & - & - & - & - & - & - & -
& - & - & - &-\tabularnewline
ESO\, 319\m G022 & 10.89 & $-$9.61 & 17.43 & 0.96 $\pm$ 0.03 & 2.40 $\pm$ 0.10 &
1.32 & 3.00 &8.85 & 8.41 $\pm$ 0.03 & 9.82 $\pm$ 0.04 & 0.98  $\pm$  0.07 & 1.44  $\pm$ 
0.12\tabularnewline
ESO\, 320\m G030 & 10.98 & $-$9.66 & 28.67 & 0.94 $\pm$ 0.03 & 2.22 $\pm$ 0.03 &
1.22 & 3.00 &8.83& 8.42 $\pm$ 0.03 & 9.90 $\pm$ 0.01 & 0.71  $\pm$  0.06 & 1.76  $\pm$ 
0.16\tabularnewline
ESO\, 440\m IG058  & 10.34 & $-$8.83 & 7.88 & 0.65  $\pm$  0.05 & 0.93  $\pm$ 
0.56 & 1.17 & 6.14& - & 8.05  $\pm$  0.04 & 9.71  $\pm$  0.10 & 0.28  $\pm$  0.04 & 0.77
 $\pm$  0.36\tabularnewline
ESO\, 267\m G030 & 11.25 & $-$9.84 & 30.32 & 1.06 $\pm$ 0.02 & 2.59 $\pm$ 0.11 &
1.25 & 2.82 & 8.64& 8.35 $\pm$ 0.03 & 9.94 $\pm$ 0.03 & 0.81  $\pm$  0.06 & 1.78  $\pm$ 
0.09\tabularnewline
IRAS\, 12116\m 5615  & 11.18 & $-$9.36 & - & 1.65 $\pm$ 0.04 & 4.25 $\pm$ 2.03
& 1.16 & 4.88 &-& 8.20 $\pm$ 0.01 & 9.86 $\pm$ 0.04 & 0.52  $\pm$  0.06 & 1.58  $\pm$ 
0.12\tabularnewline
NGC 4433  & 10.85 & $-$9.74 & 46.49 & 1.21 $\pm$ 0.04 & 2.59 $\pm$ 0.03 & 1.12 &
4.53 &8.68& 7.75 $\pm$ 0.03 & 9.94 $\pm$ 0.05 & 1.14  $\pm$  0.07 & 2.04  $\pm$ 
0.16\tabularnewline
NGC 4575 & 10.85 & $-$9.65 & 23.66 & 1.03 $\pm$ 0.03 & 1.94 $\pm$ 0.03 & 1.17 &
4.60 &8.77& 7.98 $\pm$ 0.03 & 9.95 $\pm$ 0.03 & 0.57  $\pm$  0.04 & 1.79  $\pm$ 
0.13\tabularnewline
IRAS\, 13052\m 5711 & 11.14 & $-$9.57 & 15.39 & 0.77 $\pm$ 0.03 & 1.79 $\pm$ 0.05
& 1.28 & 2.50 &8.74& 8.54 $\pm$ 0.04 & 9.89 $\pm$ 0.04 & 0.97  $\pm$  0.09 & 0.99  $\pm$ 
0.15\tabularnewline
ESO\, 221\m IG008 & 10.16 & $-$9.16 & 113.69 & 0.26 $\pm$ 0.05 & 0.88 $\pm$ 0.03
& 1.00 & 4.23 &8.36& 7.48 $\pm$ 0.03 & 9.72 $\pm$ 0.09 & 0.95  $\pm$  0.05 & 0.34  $\pm$ 
0.09\tabularnewline
ESO\, 221\m IG010 & 10.93 & $-$9.53 & 53.36 & 0.78 $\pm$ 0.04 & 1.91 $\pm$ 0.04 &
1.13 & 2.60 &8.82& 7.84 $\pm$ 0.03 & 9.99 $\pm$ 0.03 & 1.20  $\pm$  0.07 & 1.44  $\pm$ 
0.16\tabularnewline
NGC 6000 & 11.05 & $-$9.85 & 31.54 & 1.20 $\pm$ 0.01 & 2.23 $\pm$ 0.03 & 1.16 &
3.65 & 8.90&8.03 $\pm$ 0.02 & 9.91 $\pm$ 0.03 & 0.62  $\pm$  0.04 & 1.67  $\pm$ 
0.11\tabularnewline

\noalign{\smallskip}       
\hline 
\noalign{\smallskip} 

NGC\, 3690E & - & - & - & - & - & - & - & - &
- & - & - &-\tabularnewline
NGC 3690W & - & - & - & - & - & - & -  & -
& - & - & - &-\tabularnewline
IC\, 883 & - & - & - & - & - & - & - & - & - & - & - & -\tabularnewline
IRAS\, F16516\m 0948 & 11.26 & $-$9.78 & 57.11 & 1.01 $\pm$ 0.06 &
2.29 $\pm$ 0.06 & 1.10 & 3.33 &8.62& 7.46 $\pm$ 0.05 & 9.89 $\pm$ 0.08 & 0.50  $\pm$  0.06 & 0.87  $\pm$  0.13\tabularnewline
IRAS\, F17138\m1017 & 11.10 & $-$9.45 & - & 2.47 $\pm$ 0.05 &
6.08 $\pm$ 0.17 & 1.19 & 2.42&8.71 & 8.58 $\pm$ 0.09 & 10.00 $\pm$ 0.02 & - & -
\tabularnewline
IRAS\, F17578\m 0400 & 10.74 & $-$9.16 & - & - & - & -&
- & - & - & - & - & - \tabularnewline
IRAS\, 18293\m 3413 & 11.37 & $-$9.33 & 1.28 & 0.49 $\pm$ 0.02 & 1.67 $\pm$ 0.95 &
1.44 & 1.48 &-& 9.10 $\pm$ 0.01 & 9.59 $\pm$ 0.04 & - & -\tabularnewline
IRAS\, 19115\m 2124 & 11.62 & $-$9.52 & 51.27 & 1.25 $\pm$ 0.03 & 2.67 $\pm$ 0.28
& 1.16 & 2.98&8.68 & 7.84 $\pm$ 0.02 & 9.94 $\pm$ 0.03 & 0.71  $\pm$  0.05 & 1.31  $\pm$  0.14\tabularnewline
\noalign{\smallskip} 
\hline
\noalign{\smallskip} 
\multicolumn{13}{@{} p{22.5cm} @{}}{\footnotesize{\textbf{Notes.} The targets are ordered with increasing RA. Col 1: galaxy's common name; Col 2: stellar mass assuming Salpeter IMF; Col 3: specific SFR; Col 4: measured EW of H$\alpha$; Cols 5 \& 6: stellar and nebular extinction; Cols 7 \& 8: measured values of D4000 index and H$\delta$; { Col 9: estimated values of Oxygen abundances}; Cols 10 \& 11: age weighted by light and mass, respectively; Cols 12 \& 13: metallicity weighted by light and mass, respectively.}}
\end{tabular}
}
\end{table*}

\subsection{Star cluster catalogues}\label{sec:catalog}

This section briefly summarizes the methods adopted to draw the final cluster catalogues that were already used to establish the relation between the NIR brightest cluster magnitude, $M_K^{\rm brightest}$, and SFR in  \citet{2013ApJ...775L..38R}. The same catalogues are used to construct the CLFs in this work. 

We identified the star cluster candidates of the SUNBIRD galaxies following the 
selection steps presented in \citetalias{2013MNRAS.431..554R}, except that the value of the cutoff error $\sigma_m = 0.35$\,mag to include YMC candidates with slightly higher magnitude uncertainties due to the complex varying background they reside in. Had we retained $\sigma_m = 0.25$\,mag, we would have missed about 5 percent of the fainter candidates. We checked, however, and found that the choice of error cuts does not introduce photometric bias in our analysis.
FWHM versus concentration index\footnote{The concentration index $C$ is used to quantify the concentration of light in the detected object. For a S27 frame, $C = m_{\rm 1\,px} - m_{\rm 3\,px}$. The parameter is defined as  $C = m_{\rm 0.5\,px} - m_{\rm 1.5\,px}$ in
the case of S54 data.}, $C$, plots were used { to exclude contaminating sources with too narrow (stars) and too broad (background galaxies) light profiles. Given that the sample covers a wide range of luminosity distances, we defined FWHM and $C$ cutoff values that are adequate for selecting YMC candidates of each target. We note however that these selection criteria become less robust for galaxies with $D_L > 100$ Mpc where individual YMCs \citep[typical sizes of 3 -- 5 pc, e.g.][]{2021MNRAS.508.5935B} most likely appear as point-like sources. A more stringent visual inspection of the NIR images was conducted for these cases.} 

Table\,\ref{tab:YMC-num} lists the number of NIR-selected star clusters (N),  the number of YMCs above the 80\,percent completeness limit before ({ N$^{'}$}) and after ({ N$^{'}_{\rm cor}$}) applying the completeness corrections (see Section \ref{sec:comp}) to the data as well as 
the $K$-band magnitude of the brightest star clusters of the targets with unpublished CLFs. While ${\rm ESO\,221-IG008}$ with an irregular nuclear region hosts more than 400 star clusters, the number of YMCs in ESO 319\m G022, ESO 440\m IG058, NGC 3620, and IRAS 01173\p1405 only ranges between 20 and 30. These targets are not necessarily cluster-poor but their high inclination (e.g. ESO\,550\m IG050) and/or a degraded AO correction and hence a science image with a low S/N ratio (e.g. ESO\,440\m IG058) are likely to hinder the detection of YMC candidates. Since more than 95 percent of the detected clusters in ESO\,440\m IG058 and ESO\,550\m IG050 are respectively hosted by the southern and northern part of the systems, we decided to only consider the number of YMCs associated with these regions. In such cases, the values of SFR and $\Sigma_{\rm SFR}$ are derived by only considering the IR luminosity and the YMC surface areas of the southern/northern component of the two pairs of galaxies.
As for the cluster NIR luminosities, we recorded candidates that are as bright as $-18$\,mag. Although we could be looking at the most massive YMCs that form within the extreme environment of mergers such as LIRGs \citep{2019MNRAS.482.2530R}, there is also a possibility that these very bright objects are star cluster complexes, especially for $D_L > 100$\,Mpc (see Section \ref{sec:blending}). At fainter magnitude levels, we were able to detect YMCs down to $M_K \approx$\m11\,mag in this work (e.g. ESO\,221\m IG008, ESO\,428\m G023). Such a value is $\sim$\,2\,mag fainter than for the $K$-band star cluster catalogues in our pilot study. 

We note that rigorous visual inspection was done to ensure that there are no galaxy nuclei, foreground stars, and any false detections included in the final catalogues. We discuss the photometric completeness of the catalogues in Section \ref{sec:comp}.

\begin{table*}
\caption{\small The completeness levels as well as  the final number of YMC candidates for each target after imposing our selection criteria. We also tabulate { the galaxy's SFR density} and the $K$-band absolute magnitude of the brightest star cluster.} 
\label{tab:YMC-num}

\scalebox{1.}{
\begin{tabular}{lcccccccccc}   
\hline \hline
   \noalign{\smallskip}  

\multicolumn{1}{l@{\hspace{0.0cm}}}{Name} &
\multicolumn{2}{c@{\hspace{0.0cm}}}{50\,\% comp.limit } &
\multicolumn{2}{c@{\hspace{0.0cm}}}{80\,\% comp.limit } &
\multicolumn{1}{c@{\hspace{0.0cm}}}{N} &
\multicolumn{1}{c@{\hspace{0.0cm}}}{{ N$^{'}$}} &
\multicolumn{1}{c@{\hspace{0.0cm}}}{{ N$^{'}_{\rm cor}$}} &
\multicolumn{1}{c@{\hspace{0.0cm}}}{SFR} &
\multicolumn{1}{c@{\hspace{0.0cm}}}{$\Sigma_{\rm SFR}$} &
\multicolumn{1}{c@{\hspace{0.0cm}}}{$M_K^{\rm brightest}$} \\

&App mag&Abs mag&App mag&Abs mag & & & & (${\rm M_{\odot}\,yr}^{-1}$)  & (${\rm SFR/}\,{\rm kpc}^{2}$) & (mag)  \\
                ~~~~~~(1)    &   (2)        &  (3) & (4) & (5) & (6) & (7) & (8) & (9) & (10) & (11) \\     
   \noalign{\smallskip}         
\hline \hline  
\noalign{\smallskip} 
\noalign{\smallskip} 
MCG\,\m 02-01-052        & 20.4 & \m14.8 & 20.1 & \m15.1    & 41&26&27&7$^{\dagger}$ & 0.12 & $-17.21 \pm 0.13$\\
IRAS\,01173+1405     & 20.4 & \m15.1 & 20.0 & \m15.5 & 26&21&25& 73 & 3.66 &$-18.00 \pm 0.20$\\
NGC\,1134                      &21.3 & \m12.1 & 20.9 & \m12.4 & 128& 115 & 126& 12 & 0.23 &$-15.80 \pm 0.12 $\\
ESO\,550\m IG025\m N      & 21.3 & \m14.4 & 20.9 & \m14.8 &59&39&41& 30$^{\dagger}$ & 0.36 &$-17.07 \pm 0.12$\\
NGC\,1819                     &20.6 & \m13.4 & 20.2 & \m13.8 &136&114&151 & 14 & 1.10 &$-17.11 \pm 0.11 $\\
IRAS\,06164\p0311       &20.3 & \m12.8 & 20.0 & \m13.1 & 47&47&57 &11 & 0.17 &$-16.06 \pm 0.12$\\
ESO\,491\m G020              &20.5 & \m12.7 & 20.2 & \m13.0 &51&48&62& 12$^{\dagger}$ & 1.67 &$-17.47 \pm 0.19 $\\
ESO\,428\m G023              &21.5 & \m11.8 & 21.2 & \m12.0 &96&78&83  & 10 & 0.33 &$-15.51 \pm 0.15 $\\
MCG\,+02-20-003         &21.4 & \m12.8 & 21.0 & \m13.2 & 45&44&63  & 20 & 0.82 &$-16.19 \pm 0.11 $\\
IC\,2522                           &21.5 & \m11.8 & 21.1 & \m12.2 &  302&228&250&7  & 0.05 &$-15.52 \pm 0.13$\\
NGC\,3110                     &20.9 & \m13.5 & 20.6 & \m13.8&35 & 279&134&167& 0.18 &$-17.74 \pm 0.13$\\
ESO\,264\m G036            &20.5 & \m14.3 & 20.2 & \m14.6& 85&74&91&38 & 0.38 &$-17.98 \pm 0.19 $\\
ESO\,264\m G057             &21.6 & \m12.8 & 21.2 & \m13.2& 144&123&135&20 &  0.13 &$-17.15 \pm 0.21 $\\
NGC\,3508                     &20.8 & \m13.1 & 20.4 & \m13.5& 108&26&27& 14 & 0.24 &$-16.42 \pm 0.16 $\\
NGC\,3620                      &19.7 & \m12.3 & 19.5 & \m12.5&27&26&44 & 9 & 3.98 &$-15.65 \pm 0.18$\\
ESO\,319\m G022             &21.1 & \m13.2 & 20.8 & \m13.5 & 25&23&28& 19 & 0.29 &$-15.98 \pm 0.11 $\\
ESO\,320\m G030             &19.8 & \m13.6 & 19.5 & \m13.9 & 49&44&66 &21 & 2.51 &$-15.70 \pm 0.16$\\
ESO\,440\m IG058\m { S}        &21.1 & \m14.0 & 20.8 & \m14.2 & 26&21&34& { 32$^{\dagger}$} & 1.11 & $-${ 17.79 $\pm$ 0.13} \\
ESO\,267\m G030            &20.4 & \m14.2 & 20.0 & \m14.5 & 94&75&111& 26 & 0.74 &$-16.86 \pm 0.14 $\\
IRAS\,12116\m 5615       &20.0 & \m15.4 & 19.2 & \m16.2  & 45&34&38& 66 & 5.31 &$-18.42 \pm 0.14 $\\
NGC\,4433                      &20.5 & \m12.8 & 20.2 & \m13.2 &75&73&84& 13 & 0.35 &$-16.39 \pm 0.11 $\\
NGC\,4575                     &20.5 & \m12.8 & 20.2 & \m13.1 & 48&48&57& 16 & 0.91 &$-15.62 \pm 0.12 $\\
IRAS\,13052\m 5711       &21.3 & \m13.5 & 21.1 & \m13.7 & 31&31&38 & 37 & 6.54 &$-16.51 \pm 0.15 $\\
ESO\,221\m IG008             &21.8 & \m11.5 & 21.5 & \m11.9& 414&321&361& 10 & 0.20 &$-15.90 \pm 0.13$\\
ESO\,221\m IG010           &19.6 & \m13.7 & 19.4 & \m13.9 & 50&48&56  & 25  & 0.45 &$-17.19 \pm 0.13 $\\
NGC\,6000                     &20.4 & \m12.1 & 20.0 & \m12.5 &285&273&309  & 16  & 0.39 & $-16.36 \pm 0.19 $\\

\noalign{\smallskip} 
\hline 
\noalign{\smallskip} 
\multicolumn{11}{@{} p{17cm} @{}}{\footnotesize{\textbf{Notes.} Col 1: galaxy name; Cols $2 - 5$:  apparent and absolute $K$-band magnitudes of the 50 and 80 percent completeness levels, respectively. These values correspond to the middle background region where more than $\approx$\,50\,percent of the data points are below the contour level limiting that region; Col 6: number of YMCs for $\sigma \leqslant \sigma_m$; { Cols 7 \& 8: number of YMCs above the 80\,percent completeness limit before and after applying the completeness corrections to the data, respectively. The latter is used to produce the fitted CLF of the galaxy in Fig. \ref{fig:CLFs}}; Col 9: SFR based on the galaxy IR luminosity, any value marked by $\dagger$ is estimated by using the 
method described in \citet{2013ApJ...775L..38R}; { Col 10: SFR density of the host galaxy, the estimated area used to derive $\Sigma_{\rm SFR}$ is described in Section\,\ref{sec:CLF-SFRDens}};
Col 11: $K$-band absolute magnitude of the brightest star cluster. We consider only { the southern (northern)} component of the interacting system in the case of ESO\,440\m IG058 and ESO\,550\m IG050.}}
\end{tabular}
}
\end{table*}

\subsection{Completeness analysis}\label{sec:comp}

In order to estimate the completeness limits for each galaxy, we followed the same procedures as described in \citetalias{2013MNRAS.431..554R} and \citet{2019MNRAS.482.2530R}: we performed Monte Carlo (MC) completeness simulations, { which include both the detection process and the photometric analysis}, with each NIR image between 16 and 23 magnitude range in steps of 0.25\,mag. For each target, we ran the simulations within three equally-spaced regions of different background levels (while omitting the core nuclei) to derive more accurate values of the recovered completeness fractions. The simulated clusters were created from intrinsic point-source PSF models (since the clusters would be unresolved) extracted using bright and isolated stars in the fields of the actual real data frames in each relevant galaxy data set. Because  IC\,2522 and NGC\,3110 have a more complex varying diffuse background field, we defined four regions for them instead. { The latter target, along with NGC\,1134, NGC\,4575, NGC\,6000, IRAS\,01173$+$1405 and ESO\,440$-$IG058 do not have bright and isolated stars in their field.} We thus used a representative PSF model constructed from other fields but with a similar distance from their AO reference star while running their corresponding MC simulations.

Fig. \ref{fig:comp-frac} in the Appendix shows the completeness curves for all 26 main targets where the different solid lines indicate the recovered fractions from the three or four well-defined regions. The horizontal black and green dashed lines mark respectively the 50 and 80\,percent completeness levels with the  corresponding apparent and absolute magnitudes of the middle region (used as a reference) listed in Table\,\ref{tab:YMC-num}. We find that the star clusters in this region are typically 80 percent complete down to $m_{K} \sim 20.4$\,mag. This cutoff level however tends to brighten by $\approx 1 - 2$\,mag when we move toward the innermost region of a galaxy with highly variable background levels. These analyses will be considered while deriving the CLFs corrected from observational incompleteness  (see Section \ref{sec:CLFs}). 

In Sections\,\ref{sec:check-comp} and \ref{sec:CLF-rich}, we quantify the robustness of our method for computing the recovery rate of missing sources as a function of input magnitude. These will help assess the impact of completeness bias on the derived CLFs.

\section{Star cluster luminosity functions}\label{sec:CLFs}

\subsection{CLFs of the individual SUNBIRD targets}

\begin{figure*}
\centering
	{\resizebox{1.\hsize}{!}{\includegraphics[trim= 1.5cm 7cm 1.5cm 7cm]{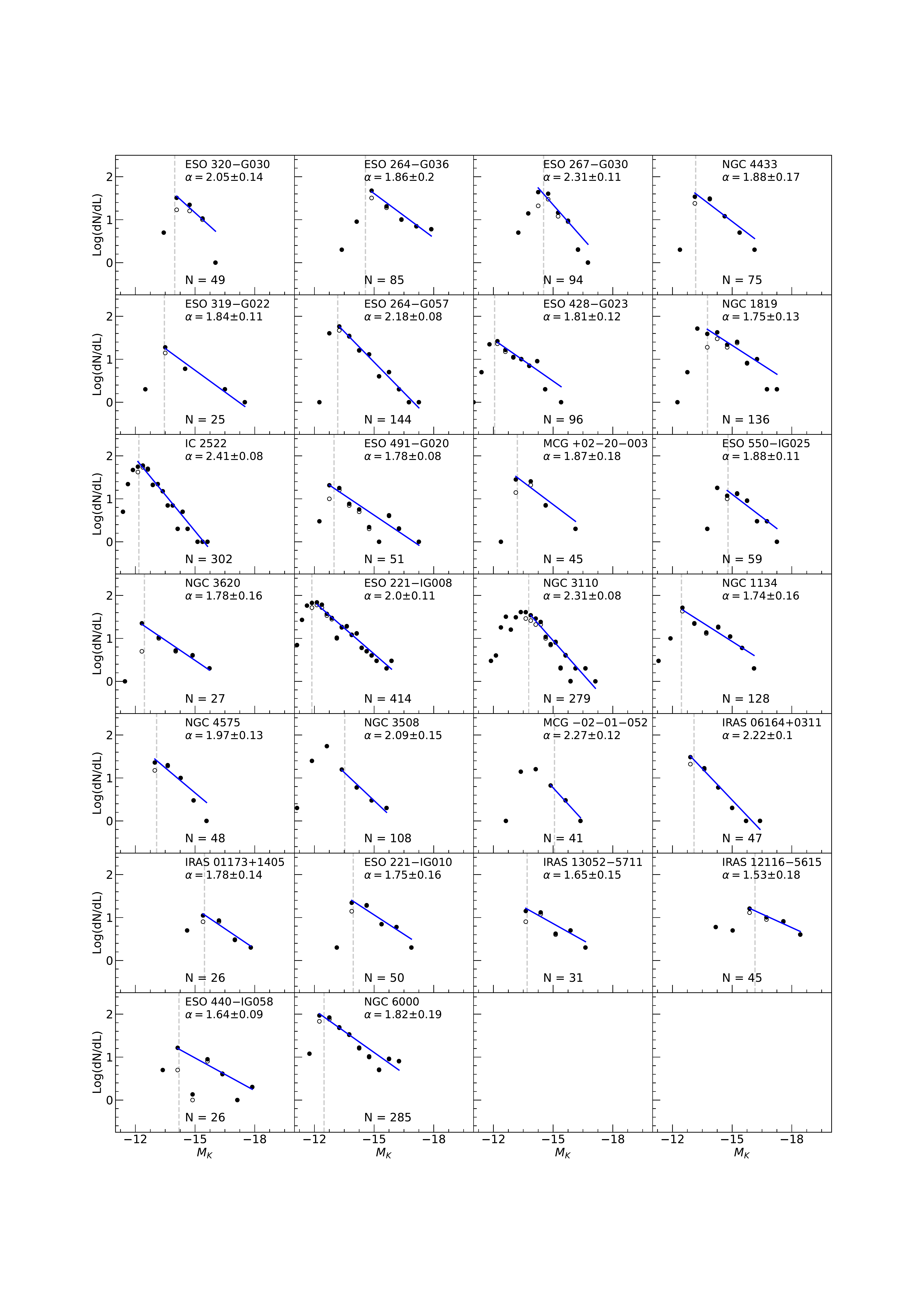}}}
    \caption{{ Individual LFs before (open circles) and after (filled circles) completeness corrections} for the YMCs hosted by the 26 SUNBIRD galaxies { using a constant magnitude binning}. For ease of comparison, the range in the y-axis is normalized to the same arbitrary number for all panels. The dashed grey line indicates the 80 percent completeness level above which a single power-law function (solid line) is fitted to the corrected data. The linear fit slightly extends beyond that vertical line whenever the value of the 80 percent level is smaller than the bin center magnitude of the last fitted bin. The derived slopes as well as the number of YMCs for each galaxy are included as insets in the CLF plots.}
    \label{fig:CLFs}
\end{figure*}

Fig. \ref{fig:CLFs} presents the binned $K$-band LFs { (open circles)} for YMC candidates hosted by the 26 galaxies in our main sample. We use a constant bin size  and then apply completeness corrections to the observed magnitudes to generate more accurate cluster luminosity distributions { (filled circles)}. For each panel, the solid line denotes the single power-law distribution, $dN/dL \propto L^{-\alpha}$, that we fit to the corrected data. Such a function is well-known to be a reasonable approximation of CLFs. The vertical bar represents the 80 percent completeness level above which we perform the fit in order to estimate the power-law slope $\alpha^{\rm cte}$ using the constant binning method. This high cutoff limit was chosen to generally coincide with the peak in the luminosity histogram and hence to ensure that most of the bins included in the fitting process are not missing star clusters. We refer to \citetalias{2013MNRAS.431..554R} for a comprehensive description of the methods used to derive the LFs and the fitted power-law slopes that are listed in the first column of Table \ref{tab:alpha}. The uncertainties in the slopes are derived from the rules of propagation of errors, considering the uncertainty in $\gamma$ from the relation $\alpha = 2.5 \gamma + 1$ where $\gamma$ is the slope of the weighted linear fit in log-log space \citep{1997ApJ...480..235E}.

The median and average of the power-law slope over the sample are respectively $\alpha_{\rm med}^{\rm cte} =1.87 \pm 0.23$ and $\alpha_{\rm aver}^{\rm cte} = 1.93 \pm 0.23$, with $\alpha$ ranging between 1.53 and 2.41. These values are consistent with the estimated CLF slopes for the 8\,LIRGs in our pilot study  where $\alpha_{\rm aver} = 1.87 \pm 0.30$ with 1.45 < $\alpha$ < 2.29. We also report in Table \ref{tab:alpha} the reduced chi-square, $\chi^{2}_{\rm red}$, of the fit to assess whether a power-law is an appropriate approximation of the CLF. { In most cases, $\chi^{2}_{\rm red} \sim 1$ which means that a single power-law fit is generally a good representation of the SUNBIRD CLFs.}

Before further discussion, we perform various analyses in Section\,\ref{sec:test-LF} to test the accuracy of our results and to identify any possible biases and uncertainties that might affect the shape of the derived CLFs. In fact, while \citet{2005ApJ...629..873M} already cautioned that the LF might be sensitive to the exact number of YMCs used in the case of a constant magnitude binning, other issues such as blending effects should also be carefully investigated \citepalias[e.g.][]{2013MNRAS.431..554R}.

\begin{table} 
\begin{scriptsize}
\caption{\small The derived power-law indices of the CLFs.}
\label{tab:alpha}
\centering
\setlength{\tabcolsep}{12pt}
\begin{tabular}{lccc}   
\hline 
   \noalign{\smallskip}  
Galaxy name  & $\alpha^{\rm cte}$ & $\alpha^{\rm var}$ & $\chi^2_{\rm red}$\\
(1)       & (2) & (3) & (4) \\   
  \noalign{\smallskip}       
\hline \hline  
\noalign{\smallskip} 

MCG\,\m02-01-052   & 2.27 $\pm$ 0.12 &  2.36 $\pm$ 0.08 & 0.04, 0.90 \\
IRAS\,01173\p1405 & 1.78 $\pm$ 0.14 & 1.90 $\pm$ 0.08 & 0.66, \m \\
NGC\,1134   & 1.74 $\pm$ 0.16 & 1.78 $\pm$ 0.10 & 2.08, 5.70\\
ESO\,550\m IG025 & 1.88 $\pm$ 0.11 & 1.63 $\pm$ 0.09  & 1.17, 7.22  \\  
NGC\,1819           & 1.75 $\pm$ 0.13  & 1.72 $\pm$ 0.10  & 2.62, 0.72\\ IRAS\,06164\p0311  & 2.22 $\pm$ 0.10 & 1.88 $\pm$ 0.09 & 0.38, 2.84 \\
ESO\,491\m G020   & 1.78 $\pm$ 0.08  &  1.74 $\pm$ 0.07 & 0.99, 0.93\\  
ESO\,428\m G023   & 1.81 $\pm$ 0.12 & 2.04  $\pm$ 0.10  & 0.92, 4.98\\
MCG\,\p02-20-003  & 1.87 $\pm$ 0.18 & 2.27 $\pm$ 0.10  & 4.81, \m\\
IC\,2522           & 2.41 $\pm$ 0.08  & 2.25 $\pm$ 0.11 & 1.29, 12.71 \\
NGC\,3110     & 2.31 $\pm$ 0.08 & 2.19 $\pm$ 0.11 & 0.82, 4.42 \\
ESO\,264\m G036   & 1.86 $\pm$ 0.20 & 1.90 $\pm$ 0.09  & 1.29, 3.38 \\
ESO\,264\m G057   & 2.18 $\pm$ 0.08 & 2.05 $\pm$ 0.09  & 0.40, 3.37 \\
NGC\,3508 	 & 2.09 $\pm$ 0.15 & 2.37 $\pm$ 0.12 & 0.33, \m \\
NGC\,3620   & 1.78 $\pm$ 0.16 & 1.66 $\pm$ 0.09 & 0.24, 5.76\\
ESO\,319\m G022   & 1.84 $\pm$ 0.11 & 1.85 $\pm$ 0.06 & 0.74, \m\\
ESO\,320\m G030   & 2.05 $\pm$ 0.14 & 2.11 $\pm$ 0.12  &  3.72, 3.32 \\
ESO\,440\m IG058 & 1.64 $\pm$ 0.09 & 1.42 $\pm$ 0.07 & 2.47, 3.92  \\
ESO\,267\m G030   & 2.31 $\pm$ 0.11 & 2.28 $\pm$ 0.10  & 2.60, 3.73 \\
IRAS\,12116\m5615  & 1.53 $\pm$ 0.18 & 1.42 $\pm$ 0.09 & 0.30, \m \\
NGC\,4433           & 1.88 $\pm$ 0.17 & 2.27 $\pm$ 0.09  & 3.07, 0.51 \\
NGC\,4575  & 1.97 $\pm$ 0.13 & 1.85 $\pm$ 0.09 & 1.98, 2.66\\
IRAS\,13052\m5711  & 1.65 $\pm$ 0.15 & 1.63 $\pm$ 0.09 & 1.09, \m \\ ESO\,221\m IG008  & 2.00 $\pm$ 0.11 & 2.02 $\pm$ 0.10 & 1.29, 10.42\\
ESO\,221\m IG010 & 1.75 $\pm$ 0.16 & 1.75 $\pm$ 0.08 & 1.29, 0.54 \\
NGC\,6000   & 1.82 $\pm$ 0.19 & 1.88 $\pm$ 0.09 & 1.96, 3.18 \\
 \noalign{\smallskip}  
   \hline
 \noalign{\smallskip} 
       { 26 Targets} \\
   Average:    & 1.93 $\pm$ 0.23  & 1.94 $\pm$ 0.27   \\  
   Median:    & 1.87  $\pm$ 0.23 &  1.88 $\pm$ 0.27  \\\\
   { 34 Targets} \\
   Average:    & 1.92 $\pm$ 0.24  & 1.93 $\pm$ 0.28   \\  
   Median:    & 1.86  $\pm$ 0.24 &  1.88 $\pm$ 0.28  \\
 \noalign{\smallskip}  
   \hline
  \noalign{\smallskip}  

\multicolumn{4}{@{} p{8cm} @{}}{\footnotesize{\textbf{Notes.} Col 1: galaxy name; Cols 2 \& 3: the indices
derived from binning with a constant and a variable bin width, respectively; Col 4: the reduced Chi Square values for the single power-law fits using the constant  and the variable binning, respectively. Note that for small data sets, there are cases where the least-square fitting fails to return a value of $\chi_{\rm red}^{2}$ or if computed, such value may not be a good representation of the goodness of the fit. The estimated average and median values of the slopes are also shown in this table, considering the main data sets of 26 targets and then all 34 SUNBIRD galaxies that have a computed CLF.}}

\end{tabular}
\end{scriptsize}
\end{table}

\subsection{Possible uncertainties and biases}\label{sec:test-LF}

\subsubsection{Choice of binning}

Even though an equal luminosity-sized binning (which we adopted in this work) is the most commonly used approach to generate the LF of star cluster systems, we also explored other methods and then compared the results. This will help confirm the authenticity of the derived LFs and subsequently the nature of the relatively shallower slopes from the fitting process. In fact, the choice of binning can affect the value of the measured power-law slope $\alpha$ as already pointed out by e.g. \citet{2005ApJ...629..873M} and \citet{2016MNRAS.462.3766C}. An artificial flattening as large as 0.3 in the binned luminosity functions might occur, especially for targets with small data sets in case of a constant binning. 

\begin{figure}
\centering
\resizebox{1.\hsize}{!}{\includegraphics{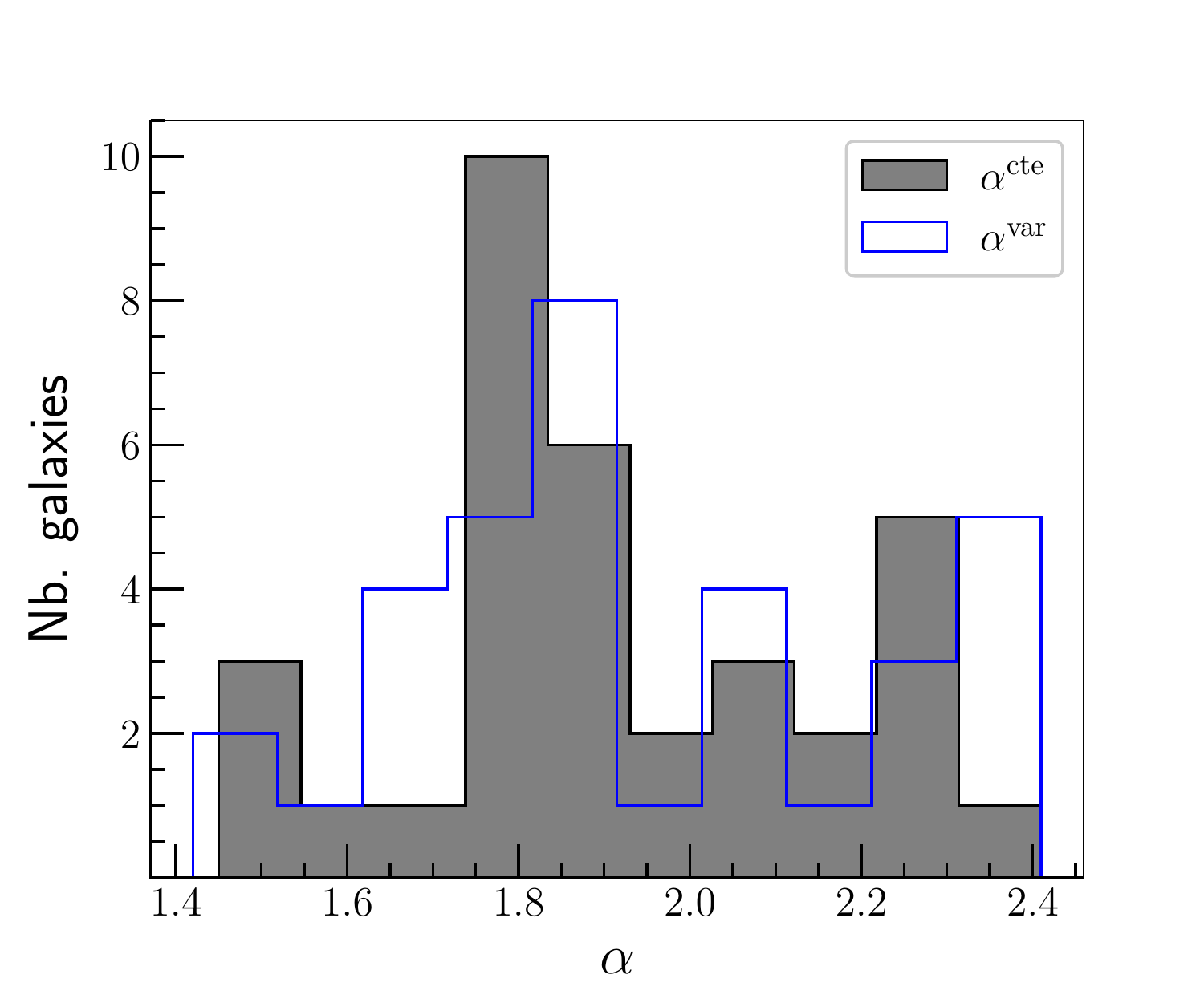}}\\
\caption{\small Distributions of the CLF power-law slopes for the 34 SUNBIRD targets, assuming a constant (cte, grey histogram) and a variable (var, blue) binning. The median values of $\alpha$ are $1.86 \pm 0.24$ and  $1.88 \pm 0.28$, respectively.}
\label{fig:compare-alpha}
\end{figure}

The grey and blue histograms of Fig.\,\ref{fig:compare-alpha} represent respectively the distribution of the derived CLF slopes from a constant (cte) and a variable (var) binning while including the whole sample (i.e. all 34 targets).
The latter method considers an equal number of clusters in each bin, of which the derived LF slopes ($\alpha^{\rm var}$) of the main targets range from 1.42 to 2.37 with a median and average value of $\alpha_{\rm med}^{\rm var} =1.88 \pm 0.27$ and $\alpha_{\rm aver}^{\rm var} = 1.94 \pm 0.27$, respectively (see third column of Table \ref{tab:alpha}). These values are consistent within their uncertainties with the CLF slopes from a constant binning. { In fact, we found that $\alpha^{\rm cte} - \alpha^{\rm var} \lesssim \pm\,0.10$ in most cases, and the one-to-one distribution between the two slopes has a general  scatter of 0.16}. They are also in agreement with the results from our pilot study in \citetalias{2013MNRAS.431..554R}: CLFs of the SUNBIRD galaxies have generally power-law slopes shallower than the canonical index of $-2$ in both studies.

We also note that additional systematic tests have shown that the choice of the luminosity-sized bin width should not statistically affect the value of $\alpha^{\rm cte}$. Changing the bin size only leads to a small scatter of $\sim$ 0.15 in the current slopes that have comparable uncertainties to this value. 

Based on these analyses, we conclude that the observed flattening in the LF slope could not be mainly caused by the choice of binning.  We consider $\alpha^{\rm cte}$ from the constant binning method in the remainder of this work, hereafter also referred to as the power-law slope $\alpha$. { It is worth mentioning that other fitting techniques such as Bayesian probabilistic modeling or maximum-likehood of cumulative functions can also be used to study the star cluster MFs/LFs (see e.g. \citealp{2017ApJ...839...78J,2018MNRAS.477.1683M,2019ApJ...872...93M,2020MNRAS.499.3267A}). Based on the comprehensive fitting analysis in Appendix B of \citet{2018MNRAS.477.1683M}, we would expect to get similar results by using these methods, especially for N\,$ > 100$.}

\subsubsection{Statistical bias and stochastic sampling}\label{sec:stat-bias}
\begin{figure}
 \begin{tabular}{c}
\resizebox{1.0\hsize}{!}{\rotatebox{0}{\includegraphics{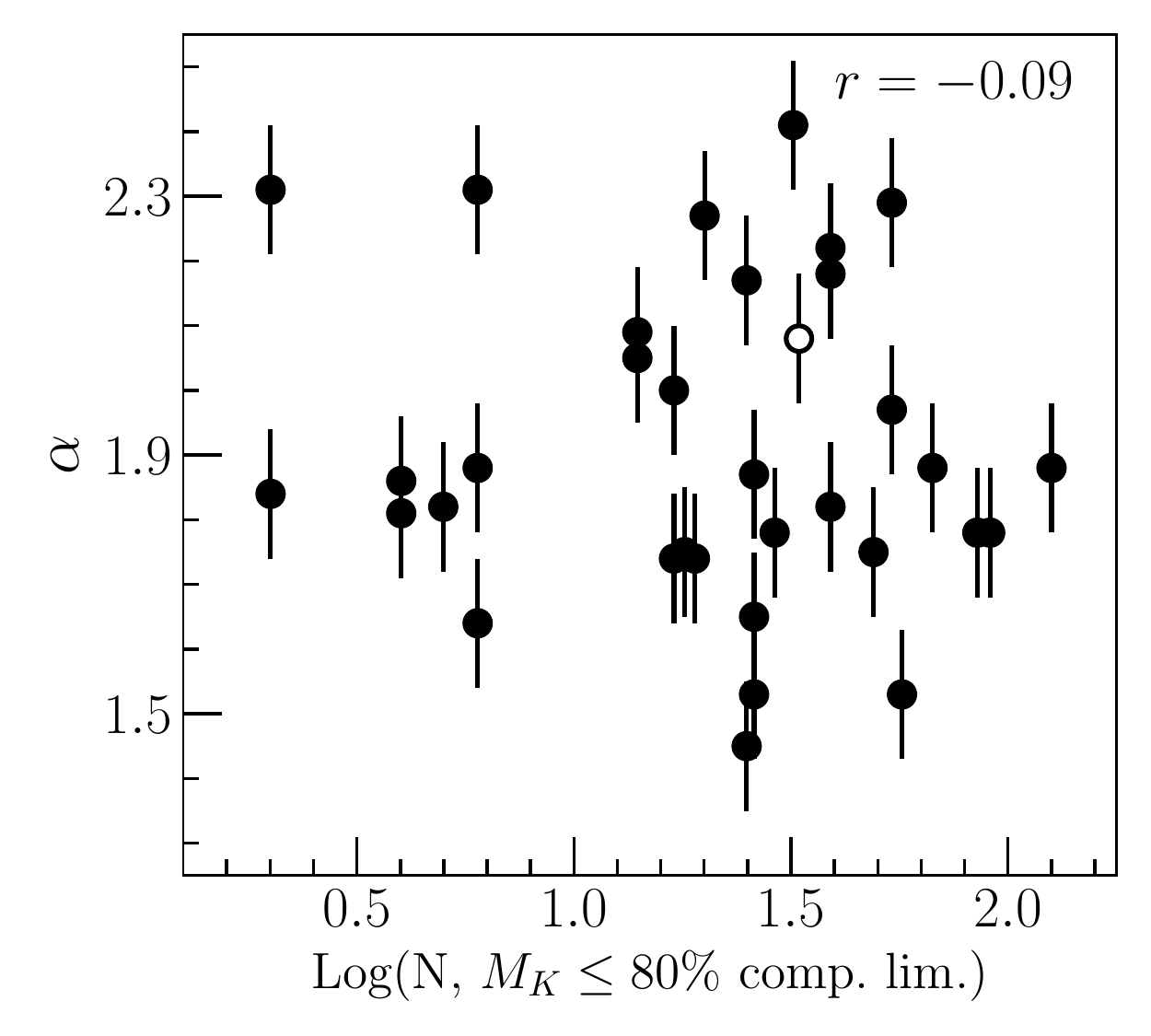}}}\\
\end{tabular}
\caption{\small The power-law slope $\alpha$ from a constant binning plotted against the number of YMCs with luminosities { brighter than the 80 percent completeness level} for each galaxy. All 34 galaxies are included in this plot with the open circle representing data points of the Bird.}
\label{fig:alpha-logN}
\end{figure}

In Fig. \ref{fig:alpha-logN}, we plot the resulting LF slope $\alpha$ against log\,N, which is the number of clusters with $K$-band luminosities brighter than { the 80 percent completeness level} for each target. By referring to the correlation coefficient\footnote{ The significance of a linear association between two variables can be measured by the Pearson correlation coefficient $r$ where $-1 < r < 1$ with $r = 0$ for no correlation and abs$(r) = 1$ denoting a perfect correlation.}, $r = -0.09$,
there is no clear trend between $\alpha$ and the parameter log\,N in the overall data.
The random scatter in $\alpha$ indicates that a flattening in the CLFs, especially for cluster-poor galaxies, cannot be caused by stochastic effects which can manifest by the presence of bright star clusters at the bright end of the LF (see e.g.\,\citealt{2016MNRAS.462.3766C}). For consistency checks, we applied other cutoff values (\m15, { \m 14.5}, \m14, and \m13.5 mag) to the data and we recorded similar trends as seen in Fig. \ref{fig:alpha-logN}, i.e. no clear correlation found between the two parameters. { Fig. 3 in \citet{2013ApJ...775L..38R} shows $M_K^{\rm brightest}$ plotted against log\,(N, $M_K \leq -15\,{\rm mag})$. Although there is a correlation between the two parameters because of size-of-sample effect, more scatter is also recorded in the magnitudes of the brightest clusters for cluster-poor galaxies.}

Nonetheless, we specifically looked at targets with a cluster-poor population, defined as N $\lesssim 30$ in this work. These galaxies are ESO 319\m G022, ESO 440\m IG058, IRAS 01173\p1405, NGC 3620, and IRAS\,13052\m 5711. They have CLF slopes ranging between $1.64\,-\,1.84$ with an average value of $1.74 \pm 0.08$. If N $< 50$, the average value becomes $1.87 \pm 0.22$ as it includes six more targets. The derived average values in both cases are relatively lower than $\alpha_{\rm aver}^{\rm cte} = 1.93 \pm 0.23$ but still consistent within uncertainties. This quick analysis is motivated by the results from \citet{2005ApJ...629..873M} 
where they have found that a spurious flattening of $\alpha$, as large as 0.3 from its original value, is highly expected for small data sets that are binned constantly due to the low-number statistics per bin.

While the overall YMC catalogues did not present any prominent statistical and stochastic effects (Fig. \ref{fig:alpha-logN}), power-law slopes of cluster-poor galaxies cannot be entirely immune from the LF binning effects. In fact, the same conclusion can be drawn from the composite CLF of cluster-poor galaxies presented in Section \ref{sec:composite-all}.

\subsubsection{PSF size of the MC completeness simulations} \label{sec:check-comp}

This section investigates the influence of the PSF size on the simulated completeness fractions and consequently, on the shape of the corrected LF and the value of its slope $\alpha$. The choice of the PSF model is essential because the data set was imaged using single conjugate AO systems, and as a result, the detected objects are expected to have different PSF sizes across the field: the closer an object is to the natural guide star, the smaller its {\tt FWHM} will be (Table \ref{tab:alpha-psf}).

\begin{figure}
\centering
\resizebox{1.\hsize}{!}{\includegraphics[trim= 0cm 0cm 0cm 0.3cm]{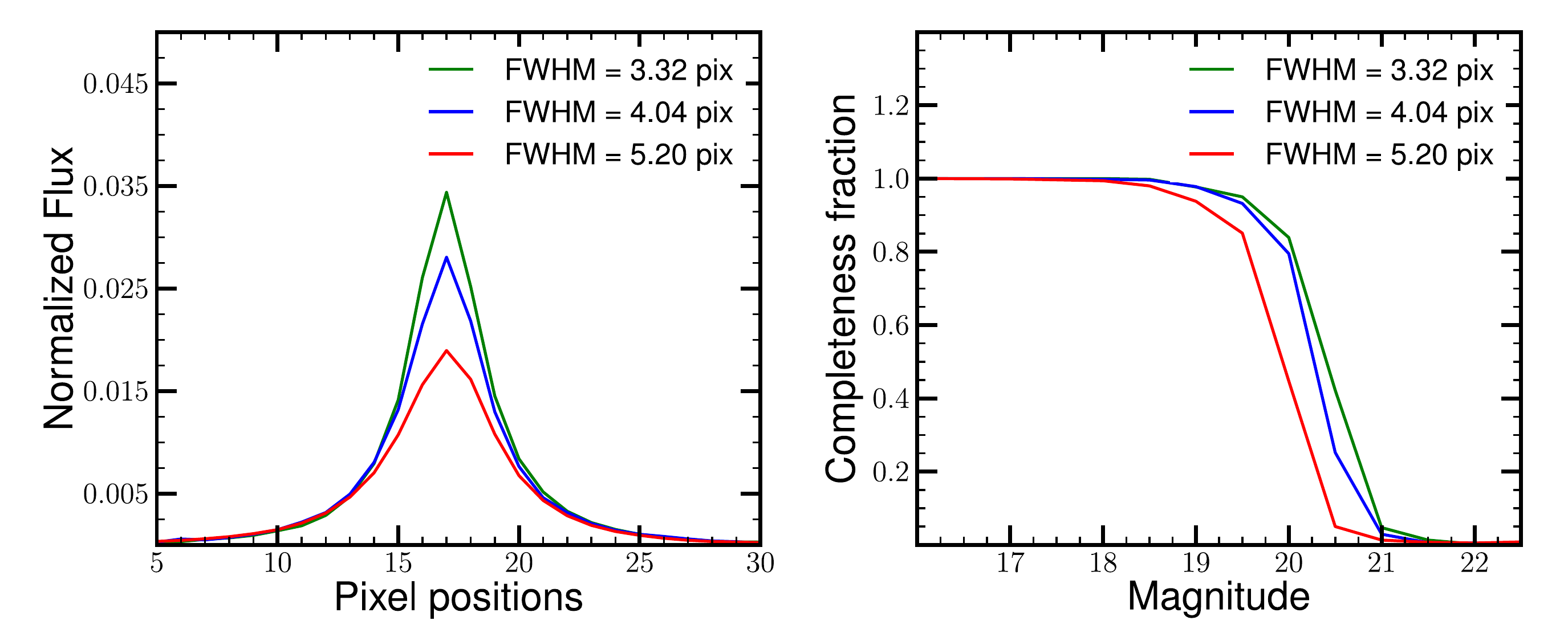}}\\
\resizebox{.98\hsize}{!}{\rotatebox{0}{\includegraphics[trim= 0cm 0cm 0cm .6cm]{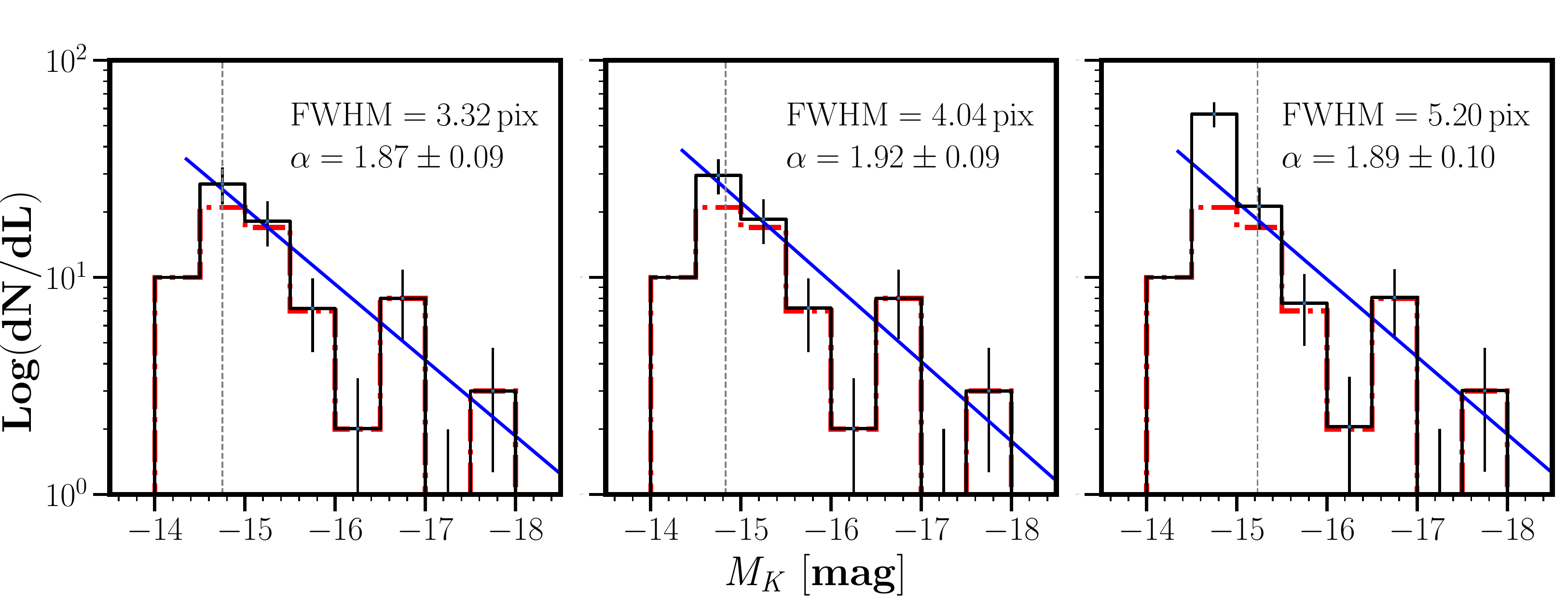}}}
\caption{\small {\it Upper left:} radial profile of the PSF used to create artificial PSF star. {\it Upper right:} Completeness fractions as a function of the PSF used. {\it Lower panels:} $K$-band CLFs of ESO\,264\m G036 corrected by different completeness fractions. The black solid line is the completeness-corrected LF while the dash-dotted red line is the original one. The blue solid line represents the single power-law fit to the CLF. The dashed grey line marks the 80\,percent completeness level.}
\label{fig:lf-psf}
\end{figure}

We performed a varying PSF test using point sources in the field of ESO\,264\m G036. Three bright and isolated stars scattered over the field
were selected to represent different PSFs. The upper left panel of Fig.\,\ref{fig:lf-psf} depicts the radial profiles of the sharp (green and blue) and the extended (red) PSF models. We then generated three sets of completeness fractions at each magnitude level based on these representative models. The upper right panel of Fig.\,\ref{fig:lf-psf} indicates that the trends of the completeness curves are consistent with the varying size of the input models: simulations performed with a wider PSF (red curve) record lower completeness of detections as we go towards fainter magnitudes compared to the ones that use a narrower PSF (the other curves).

If we fit the corrected LFs until their respective 80\,percent completeness levels
(as listed in Table \ref{tab:alpha-psf}), { we get power-law indices ranging between $1.87 - 1.92$. These values are consistent with $1.86 \pm 0.20$, which is the value of the slope recorded in Table \ref{tab:alpha} for ESO\,264\m G036.} There are no significant changes in the shapes of the LFs, except the second-last magnitude bin of the LF that was corrected using completeness fractions computed from an extended PSF model (see the bottom panels of Fig.\,\ref{fig:lf-psf}). This particular bin is however already below the 80\,percent completeness level and hence, would be excluded from our analysis. 

These results show that the location of the selected PSF stars, either close or distant from the NGS, does not introduce a significant bias toward the shape and the slope of the derived LF. 
Using a single PSF model is therefore a reasonable approximation to generate the simulated completeness fractions throughout the host galaxy field. 

We refer to Section \ref{sec:CLF-rich} which presents the CLFs for cluster-rich galaxies but also investigates further the dependence of the computed completeness fractions on the defined regions used to run the MC simulations.

\begin{table} 
\centering
\begin{scriptsize}
\caption{\small Properties of the PSF stars in the field of ESO\,264\m G036 and the resulting LF slopes.}
\label{tab:alpha-psf}
\centering
\begin{tabular}{ccccc}   
\hline
   \noalign{\smallskip}  

PSF/FWHM  & $m_{\rm PSF}$ & AO-dist & Comp.lims & $\alpha$ \\
(pix)     & (mag) & (arcsec) & (50, 80)\% (mag) & \\
(1)       & (2)   & (3) & (4) & (5) \\   
  \noalign{\smallskip}       
\hline \hline  
\noalign{\smallskip} 
3.32  & 12.4 & 24.3 & \m14.4, \m14.8 & 1.87 $\pm$ 0.09   \\
4.04  & 13.7 & 36.4 & \m14.5, \m14.8 & 1.92 $\pm$ 0.09   \\
5.20  & 11.1 & 50.9 & \m14.9, \m15.2 & 1.89 $\pm$ 0.10  \\
 \noalign{\smallskip}  
   \hline
\multicolumn{5}{@{} p{7cm} @{}}{\footnotesize{\textbf{Notes.} Cols 1 \& 2: FWHM and apparent magnitude of the PSF star; Col 3: its distance to the AO-NGS; Col 4: absolute magnitudes of the 50 and 80 percent completeness levels, respectively; Col 5: value of the power-law index from the fitted LF.}}
\end{tabular}
\end{scriptsize}
\end{table}

\subsubsection{Resolution bias}\label{sec:blending} 
Investigating the effect of spatial resolution on the CLFs for a SUNBIRD subsample is among our key results in \citetalias{2013MNRAS.431..554R}. Various methods such as Monte Carlo-based blending simulations in LFs as well as SC analysis in a redshifted Antennae Galaxies were conducted and we have shown that the resulting power-law slopes of the LIRGs with distance $D_L < 100$\,Mpc should only flatten by $\sim$\,0.15 at most because of resolution bias. 

\begin{figure*}
\centering
\begin{tabular}{c}
{\resizebox{1.0\hsize}{!}{\includegraphics{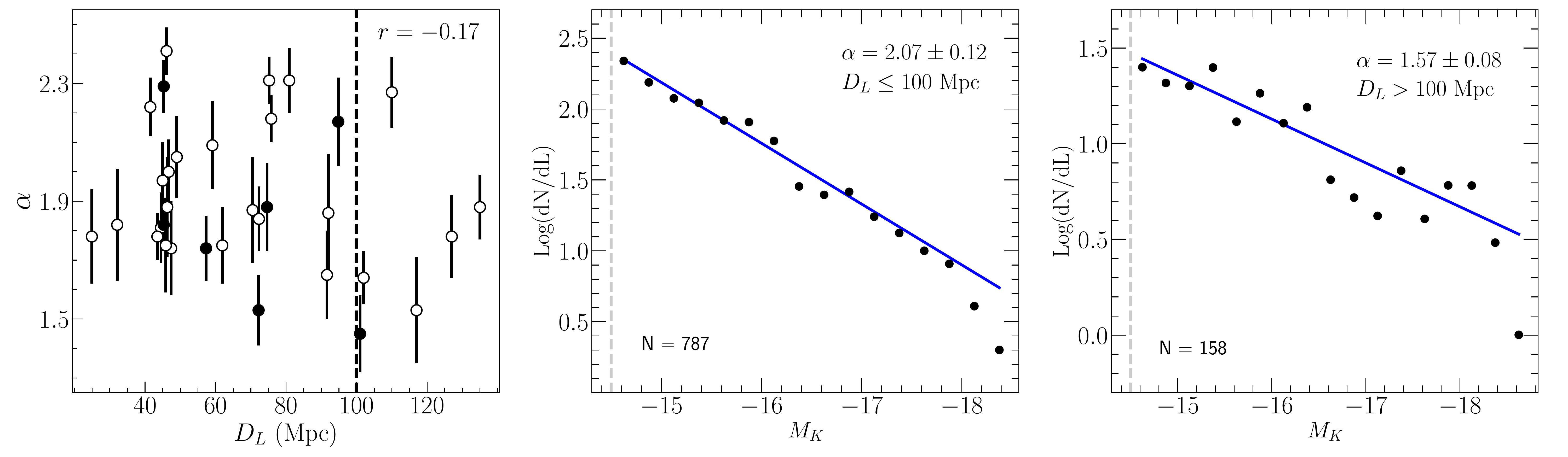}}}
\end{tabular}
\caption{\small  {\it Left:} $\alpha$ plotted against the luminosity distance of the host galaxy. We do not find any systematic trend between the two parameters. Data points associated with the main targets in this work are labelled as open circles, whereas filled circles correspond to the LIRGs studied in \citetalias{2013MNRAS.431..554R}. The dashed line marks distance of 100 Mpc. {\it Middle \& Right: } Completeness-corrected CLFs of two subsamples generated from our observational data but segregated by a distance cutoff: $\alpha = 2.07$ for $D_{L} \leq 100\,{\rm Mpc}$ and $\alpha = 1.57$ for $D_L  > 100\,{\rm Mpc}$. Data set from the Bird is excluded to avoid bias in the analysis.}
\label{fig:alpha-DL}
\end{figure*}

In spite of these findings and the use of a small radius of 2 or 3 pixels ($\sim$\,0.1 arcsec) for aperture photometry (a radius which corresponds to a physical size of $\sim 12 - 60$\,pc at the distance of our targets), we also run additional tests as a further check to the significance of spatial resolution bias on the overall SUNBIRD CLFs, excluding the Bird. We thus draw the following plots:

\begin{enumerate}

\item the power-law slope $\alpha$ plotted against the luminosity distance of the sample in the left panel of Fig. \ref{fig:alpha-DL}. The open circles correspond to the main targets with their individual CLFs shown in Fig.\,\ref{fig:CLFs}, whereas the filled circles represent the old data from our pilot study. The data points are randomly distributed, especially for targets with $D_L \leq 100$ Mpc (dashed line) where the average value of the individual slopes is $\alpha_{\rm aver} = 1.94 \pm 0.23$. For the most distant targets, i.e. $D_L > 100$ Mpc, $ 1.45 < \alpha < 2.27$ and $\alpha_{\rm aver}= 1.76 \pm 0.27$. 

\item the composite CLFs for targets out to distances $D_L \leq 100$ Mpc and then $D_L > 100$ Mpc in the middle and right panels of Fig.\,\ref{fig:alpha-DL}, respectively. We apply a single power-law fit to the combined { completeness-corrected} data down to a cutoff level of $-14.5$ mag.  The derived slopes are $2.07 \pm 0.12$ and $1.57 \pm 0.08$, respectively. If we split further the SC catalogues from the relatively less distant targets into $D_L \leq 60$ Mpc and $60 < D_L {\rm (Mpc)}\leq 100$, we get CLF slopes of $2.15 \pm 0.07$ and $2.02 \pm 0.12$, respectively. We note that the cutoff limit of $-14.5$ mag will be used throughout the paper as a value giving more than 50 percent completeness in most galaxies, except for MCG\m02-01-052, IRAS 01173\p1405, and IRAS 12116\m5615 (refer to Section \ref{sec:comp} and Table \ref{tab:YMC-num}). We have checked that going down to this limit does not affect the slope recovered from the power-law fitting. In fact, while considering a brighter cutoff of $-15$ mag, the values of $\alpha$ respectively become $2.12 \pm 0.12$  ($D_L \leq 100$ Mpc) and $1.64 \pm 0.09$ ($D_L > 100$ Mpc). These are consistent with the derived values of $\alpha$ for a limit of $-14.5$ mag.

\end{enumerate}

\begin{figure*}
\centering
\begin{tabular}{c}
{\resizebox{1.\hsize}{!}{\includegraphics{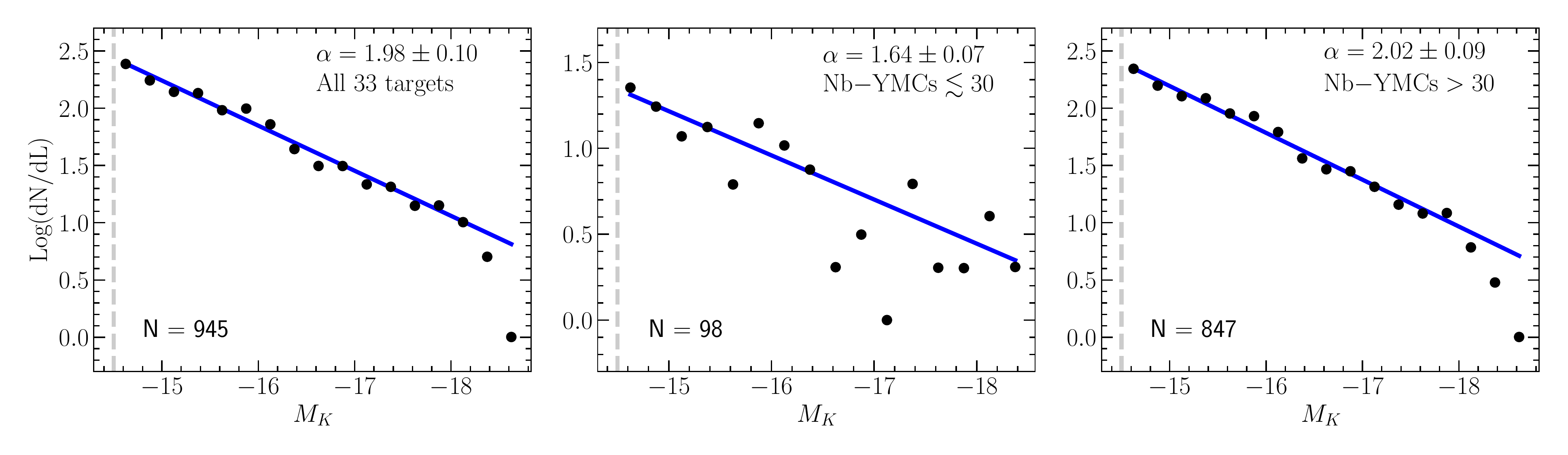}}}\\
\end{tabular}
\caption{\small Completeness-corrected composite CLFs of the 33 SUNBIRD targets { {\em (left panel)}, targets with N $\lesssim 30$ {\em (middle)} and then N $ > 30$ {\em (right)}.} The dashed grey line marks a completeness level of \m14.5 mag. Other labels are the same as in Fig. \ref{fig:CLFs}.}
\label{fig:CLF-all}
\end{figure*}

The wide scatter of the data in the left panel of Fig. \ref{fig:alpha-DL} and the indices that are similar to the canonical slope $\alpha = 2$ for $D_L \leq 100$ Mpc in the middle panel indicate that spatial resolution effect does not have a strong influence on the derived power-law slopes of the less distant targets.  These results are in good agreement with the blending analyses reported in \citetalias{2013MNRAS.431..554R}: at distances below 100\,Mpc, the effect of spatial resolution should not be a major issue and cannot have sole responsibility for any small value of $\alpha$. Beyond that distance, the effect is likely more prominent and by artificially flattening the CLF as more star cluster complexes populate the bright magnitude bins. There are 6 SUNBIRD targets that have distances $100 < D_L {\rm (Mpc)}\leq 135$. 

\subsubsection{Extinction}

The star cluster catalogues presented in Section \ref{sec:catalog} are not corrected for foreground galactic extinction which is negligible in the NIR regime. And we could not estimate the extinction of each individual YMC since our current analysis is based on observations with a single filter. The $K$-band magnitudes used to derive the CLFs in Fig.\,\ref{fig:CLFs} are thus not de-reddened, yet they are already associated with shallower power-law slopes. The extinction effect is likely more significant to the YMCs residing in the nuclear regions of the SUNBIRD galaxies. As an illustration,  \citet{2019MNRAS.482.2530R} derived a visual extinction $A_{\rm v}$ between 2.5 to 4.5 mag for the YMCs hosted by the inner regions of Arp 299 while the ones outside of the highly obscured regions have an extinction $A_{\rm v} \lesssim 0.8$ mag. \citet{2020ApJ...888...92L} indicated that a significant extinction effect would further decreases the measured power-law slope of local LIRGs. The flattening arises because more data points will populate the bright magnitude bins.

Finally, the multi-band study of the YMC population in Arp 299 also revealed that regardless of the optical filter used ($U$, $B$ and $I$), the fitted CLF of this interacting LIRG always returns a shallower slope ranging between $1.61 - 1.81$ in that wavelength regime \citep{2019MNRAS.482.2530R}. Based on these findings and due to the lack of NIR CLF works in the literature to compare directly with our results, it is thus still reasonable to compare the power-law slopes of our SUNBIRD $K$-band CLFs to those of { nearby galaxies with less intense SF activity} that are primarily based on optical observations.

\subsection{CLFs of composite "Supergalaxies"}
In this section, we create LFs from composite "supergalaxies" to increase the number statistics of the YMCs hosted by the SUNBIRD sample. { All presented CLFs are drawn using a constant magnitude binning and corrected for observational incompleteness.} Such analyses will help us explore the nature of the LF and its measured power-law slope. To avoid bias, we do not include data points from the Bird  where the cluster magnitudes are significantly different to the rest of the catalogue \citepalias{2013MNRAS.431..554R}. 

\subsubsection{Considering all SUNBIRD galaxies}\label{sec:composite-all}

{ The left panel of} Fig. \ref{fig:CLF-all} shows a { completeness-corrected} composite LF from the combined star clusters of 33 galaxies. We apply a power-law fit (solid line) to the CLF down to a magnitude limit $M_K = -14.5$\,mag where { 80 percent of the SUNBIRD data should be complete}. We measure a bright end slope of $\alpha = 1.98 \pm 0.10$ which is consistent with the average and median slopes of individual LFs presented in Section\,\ref{sec:CLFs}. The exponent of the composite LF spans between 1.98 and 2.02 while varying the magnitude limit from \m15 to \m14\,mag. Table\,\ref{tab:alpha-composite} lists the number of YMCs composing various luminosity-limited combined catalogues as well as the resulting slopes as a function of the cutoff magnitude.

{ Fig. 9 also shows the composite CLFs of two supergalaxies with N $\lesssim 30$ (middle panel) and N $> 30$ (right). By fitting a power-law to the data down to $M_K = -14.5$ mag, the computed values of $\alpha$ are $1.64 \pm 0.07$ and $2.02 \pm 0.09$. These values become $1.77 \pm 0.08$ and $2.05 \pm 0.11$, if we change the cutoff number of YMCs from 30 to 50 when splitting the SUNBIRD sample. The value of $\alpha$ for the subsample with N > 30 (N > 50) is also consistent with the average and median slopes of individual CLFs reported in Table \ref{tab:alpha} . However, statistical bias which is already discussed in Section \ref{sec:stat-bias} appears to affect the composite CLF of cluster-poor galaxies and hence a smaller value of the corresponding slope.}

The composite LFs { in the left and right-hand panels} also reveal a possible break { near a magnitude $M_K \sim -18$\,mag}. This downturn is also { marginally seen in the other composite CLFs in Figs. \ref{fig:alpha-DL}, \ref{fig:LFs-SFR}, \ref{fig:LFs-sSFR}, and \ref{fig:LFs-SFRDens}. Such a feature could not be related to observational incompleteness, since it is associated to the brightest clusters in the sample. It could be related to the high-mass break observed in some galaxies in the nearby universe (see e.g. \citealp{2019ARA&A..57..227K}). However, linking the cluster luminosity to the mass distribution is not trivial, due to the lack of a direct  one-to-one correspondence between them (see e.g. \citealp{2009A&A...494..539L}), and is beyond the scope of the current work.} 

\begin{table} 
\begin{scriptsize}
\caption{\small The values of $\alpha$ with varying magnitude limits of the composite CLF.}
\label{tab:alpha-composite}
\setlength{\tabcolsep}{10pt}
\begin{center}
\begin{tabular}{cccc}   
\hline 
 \noalign{\smallskip}  

Mag.cutoff  & N & $\alpha$ & $\chi^2_{\rm red}$\\
(1)       & (2) & (3) & (4) \\   
  \noalign{\smallskip}       
\hline \hline  
\noalign{\smallskip} 
$-15.0$  & 646  & 2.02 $\pm$ 0.10  & 1.44 \\
$-14.5$  & 945  & 1.98 $\pm$ 0.10  & 1.41 \\
$-14.0$  & 1327 & 2.01 $\pm$ 0.10  & 1.35 \\

 \noalign{\smallskip}  
   \hline
   \noalign{\smallskip}
\multicolumn{4}{@{} p{6cm} @{}}{\footnotesize{\textbf{Notes.} Col 1: the magnitude cutoff level used to draw the composite catalogue; Col 2: the number of YMCs in that catalogue; Cols 3 \& 4: the CLF power-law slope and the corresponding value of $\chi^2_{\rm red}$.}}
\end{tabular}
\end{center}
\end{scriptsize}
\end{table}

\subsubsection{Sample split as a function of SFR}\label{sec:CLF-SFR}

\begin{figure*}
\centering
\resizebox{1.\hsize}{!}{\includegraphics[trim= 0cm 0.4cm 0cm 0.0cm]{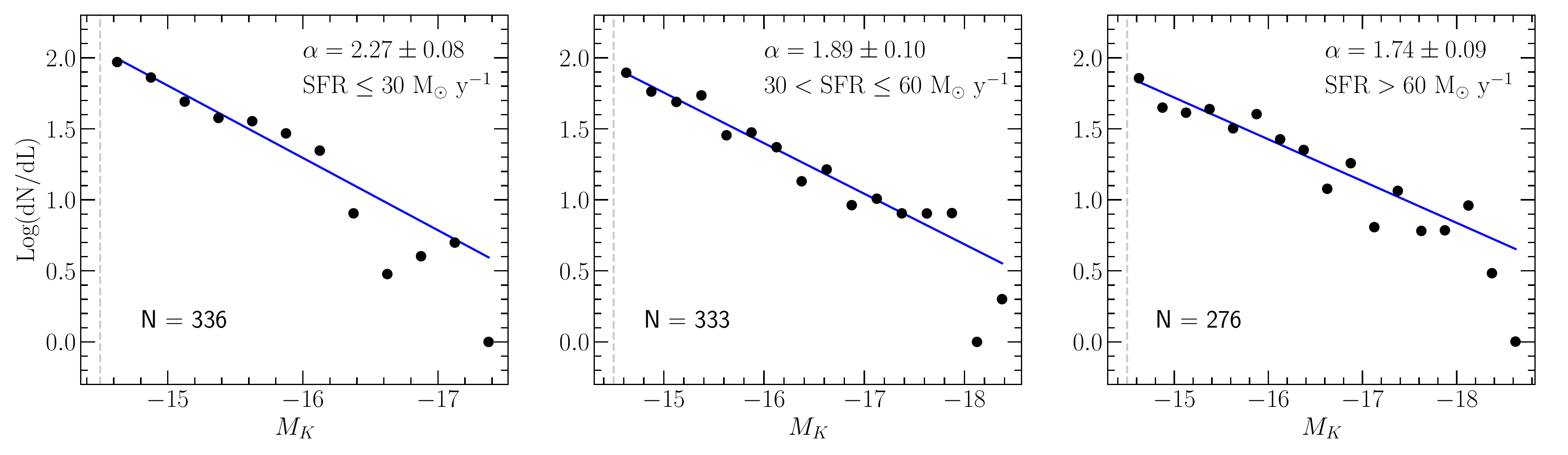}}\\
\caption{\small { Completeness-corrected} composite CLFs with the individual star cluster catalogues grouped as a function of the host galaxy's SFR: $\alpha = 2.27$ for SFR $\leq 30\,{\rm M_{\odot}\,yr^{-1}}$, $\alpha = 1.89$ for $30 < {\rm SFR  \leq 60\,M_{\odot}\,yr^{-1}}$, and $\alpha = 1.74$ for SFR $ > 60\,{\rm M_{\odot}\,yr^{-1}}$. Labels are the same as in Fig. \ref{fig:CLFs}.}
\label{fig:LFs-SFR}
\end{figure*}

\begin{figure*}
\centering
\resizebox{0.8\hsize}{!}{\includegraphics[trim= 0cm 0.0cm 0cm 0cm]{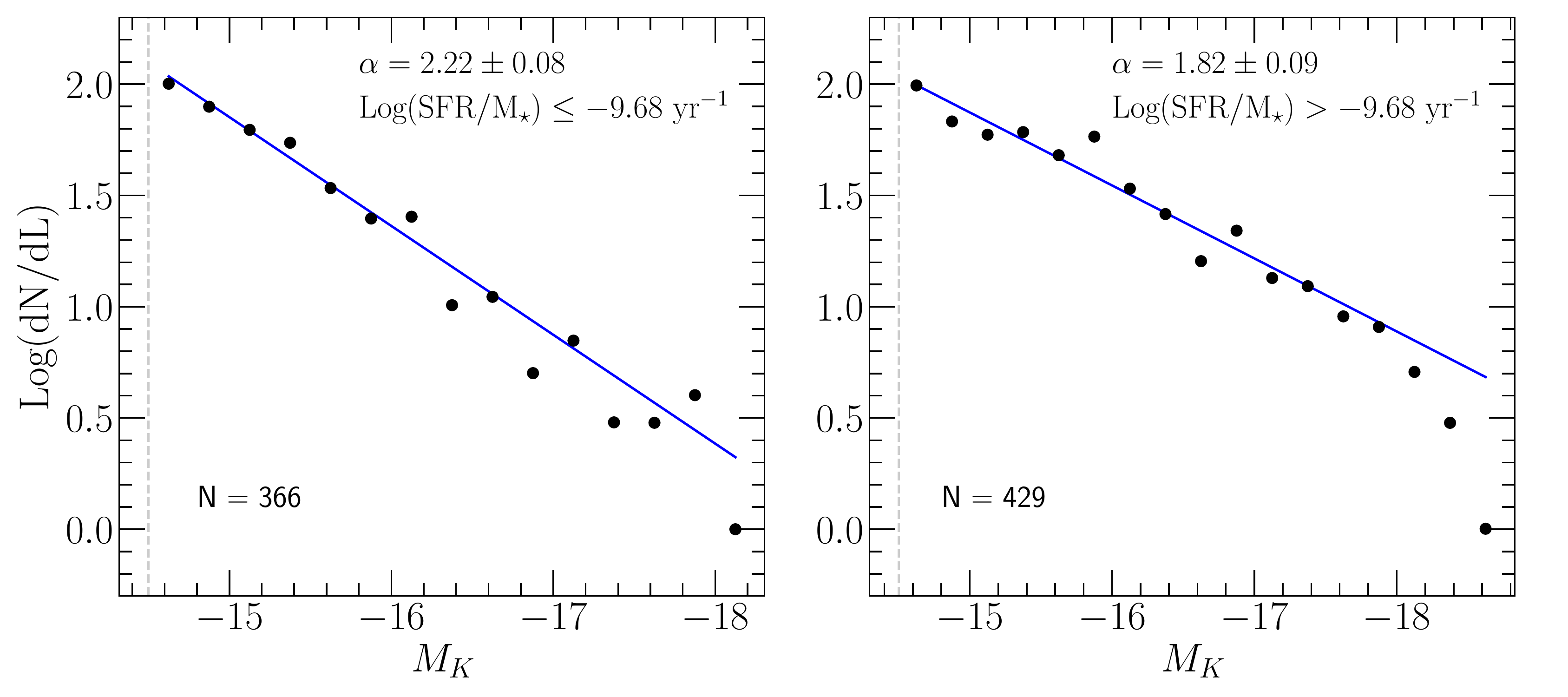}}\\
\caption{\small { Completeness-corrected} composite CLFs with the individual star cluster catalogues grouped as a function of the host galaxy's specific SFR: $\alpha = 2.22$ for Log\,(SFR/M$_{\star}) \leq -9.68\,{\rm yr^{-1}}$ and $\alpha = 1.82$ for Log\,(SFR/M$_{\star}) > -9.68\,{\rm yr^{-1}}$. Labels are the same as in Fig. \ref{fig:CLFs}.}
\label{fig:LFs-sSFR}
\end{figure*}

\begin{figure*}
\centering
\resizebox{0.8\hsize}{!}{\includegraphics[trim= 0cm 0.0cm 0cm 0cm]{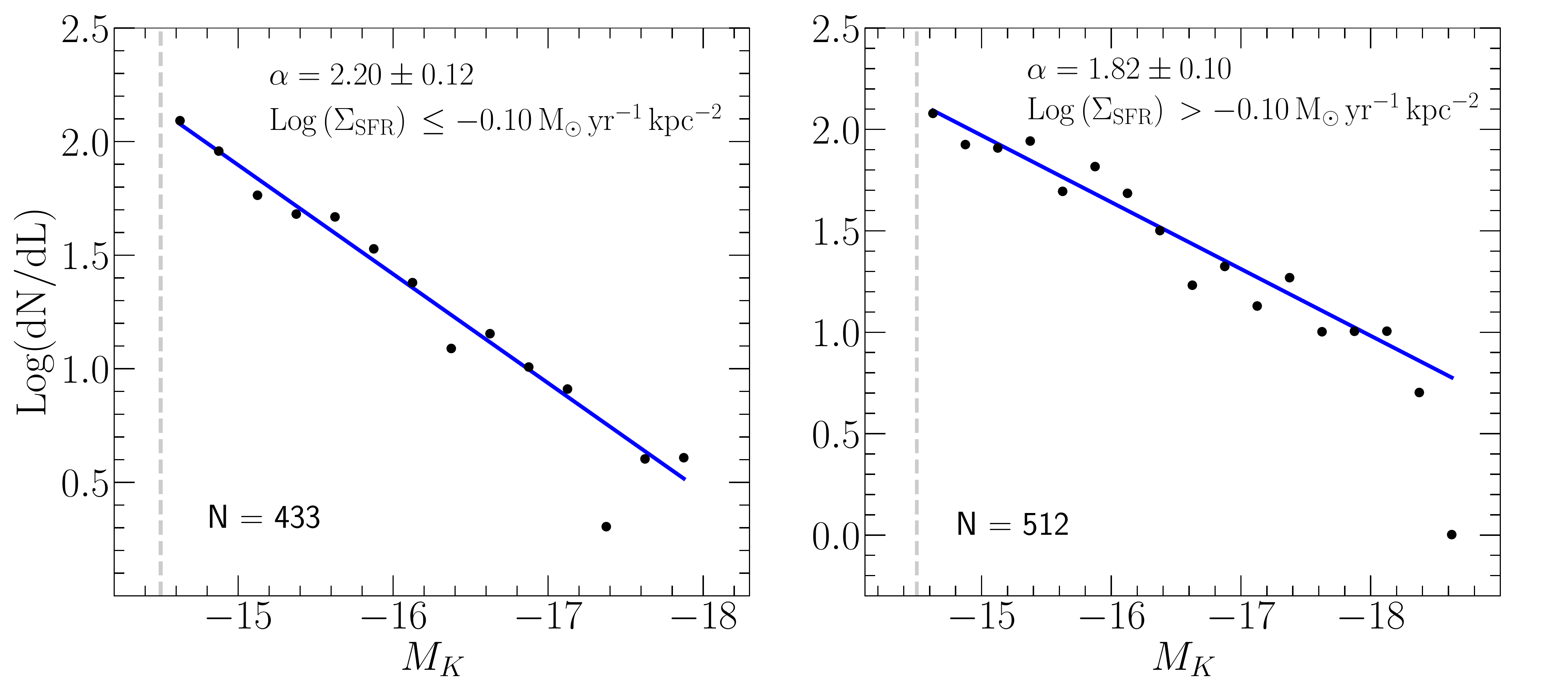}}\\
\caption{\small { Completeness-corrected composite CLFs with the individual star cluster catalogues grouped as a function of the host galaxy's SFR density: $\alpha = 2.20$ for Log\,($\Sigma_{\rm SFR}) \leq -0.10\,{\rm M_{\odot}\,yr^{-1}\,kpc^{-2}}$ and $\alpha = 1.82$ for Log\,($\Sigma_{\rm SFR}) > -0.10\,{\rm M_{\odot}\,yr^{-1}\,kpc^{-2}}$. Labels are the same as in Fig. \ref{fig:CLFs}.}}
\label{fig:LFs-SFRDens}
\end{figure*}

To see if there is a correlation between the slope $\alpha$ and the SFR of the host galaxy, we construct composite LFs of SUNBIRD subsamples that were generated as a function of SFR. Fig. \ref{fig:LFs-SFR} plots LFs from three composite supergalaxies with ${\rm SFR} \leq 30\,{\rm M_{\odot}\,yr^{-1}}$ (left panel), then $30 < {\rm SFR\,(M_{\odot}\,yr^{-1})} \leq 60$ (middle), and ${\rm SFR} > 60\,{\rm M_{\odot}\,yr^{-1}}$ (right). These cutoff values of SFR were chosen given that the average SFR is $\sim 30\,{\rm M_{\odot}\,yr^{-1}}$ but also to produce composite catalogues that have more or less a similar number of YMCs. 

The data are fitted to a power-law function down to $M_K = -14.5$\,mag as already performed previously. The derived values of $\alpha$ are $2.27 \pm 0.08$, $1.89 \pm 0.10$ and $1.74 \pm 0.09$, respectively. We get  $\alpha = 1.83 \pm 0.10$ if we combine all galaxies with ${\rm SFR} > 30\,{\rm M_{\odot}\,yr^{-1}}$ into a single subsample and $\alpha = 2.11 \pm 0.09$ for a composite of ${\rm SFR} \leq 60\,{\rm M_{\odot}\,yr^{-1}}$. The value of the power-law slope appears to decrease with an increasing SFR: steeper for galaxies with less intense SF activity in comparison to those with high SFRs. In Section \ref{sec:search}, we also search for any correlation between the individual CLF slope and the SFR to better investigate the observed trend prior to any physical interpretation. 

\subsubsection{Sample split as a function of sSFR}\label{sec:CLF-sSFR}
Fig. \ref{fig:LFs-sSFR} also shows two composite CLFs that were derived as a function of the host galaxy's sSFR(=SFR/M$_{\star}$):  log(SFR/M$_{\star}) \leq -9.68\,{\rm yr^{-1}}$ (left panel) and log(SFR/M$_{\star}) > -9.68\,{\rm yr^{-1}}$ (right). This cutoff value was chosen to ensure that there is a similar number of YMCs in the derived catalogues, each coming from the SC population of 13 galaxies. We note that \citet{2018PhDT.......244R} estimated sSFR of 26 out of the 34 SUNBIRD targets used for correlation searches with $\alpha$.

By fitting a power-law function to the composite CLFs, we get $\alpha = 2.22 \pm 0.08$ and $\alpha = 1.82 \pm 0.09$, respectively. The power-law slope  decreases with an increasing specific SFR. We will also explore further this trend in Section \ref{sec:search}.

\subsubsection{Sample split as a function of $\Sigma_{\rm SFR}$}\label{sec:CLF-SFRDens}

{ Finally, we present in Fig. \ref{fig:LFs-SFRDens} composite CLFs of two supergalaxies that were constructed with respect to the SFR density of the host galaxy: log\,($\Sigma_{\rm SFR}) \leq -0.10\,{\rm M_{\odot}\,yr^{-1}\,kpc^{-2}}$(left panel) and  log\,($\Sigma_{\rm SFR}) > -0.10\,{\rm M_{\odot}\,yr^{-1}\,kpc^{-2}}$ (right). We chose this cutoff given that it is the average value of the SFR density that spans between $\sim  0.05 - 7.25~{\rm M_{\odot}\,yr^{-1}\,kpc^{-2}}$.

The area used to derive $\Sigma_{\rm SFR}$ is estimated by considering the projected area where the YMCs are located in each galaxy. This area usually corresponds to a region starting at a level of 3$\sigma$ above the sky background. We list the values of $\Sigma_{\rm SFR}$ in Table\,\ref{tab:YMC-num}.

We fit the completeness-corrected CLF to a power-law function and we get $\alpha = 2.20 \pm 0.12$ and $\alpha = 1.82 \pm 0.10$, respectively. The value of $\alpha$ decreases by $\sim$\,0.4 with an increasing SFR density. A further analysis is conducted in Section \ref{sec:search} to better understand the nature of the observed trend.}

\subsection{CLFs on sub-galactic scales}\label{sec:CLF-rich}

\begin{figure*}
\centering
{\resizebox{.9\hsize}{!}{\includegraphics[trim= 0cm 0cm 0cm 0cm, clip]{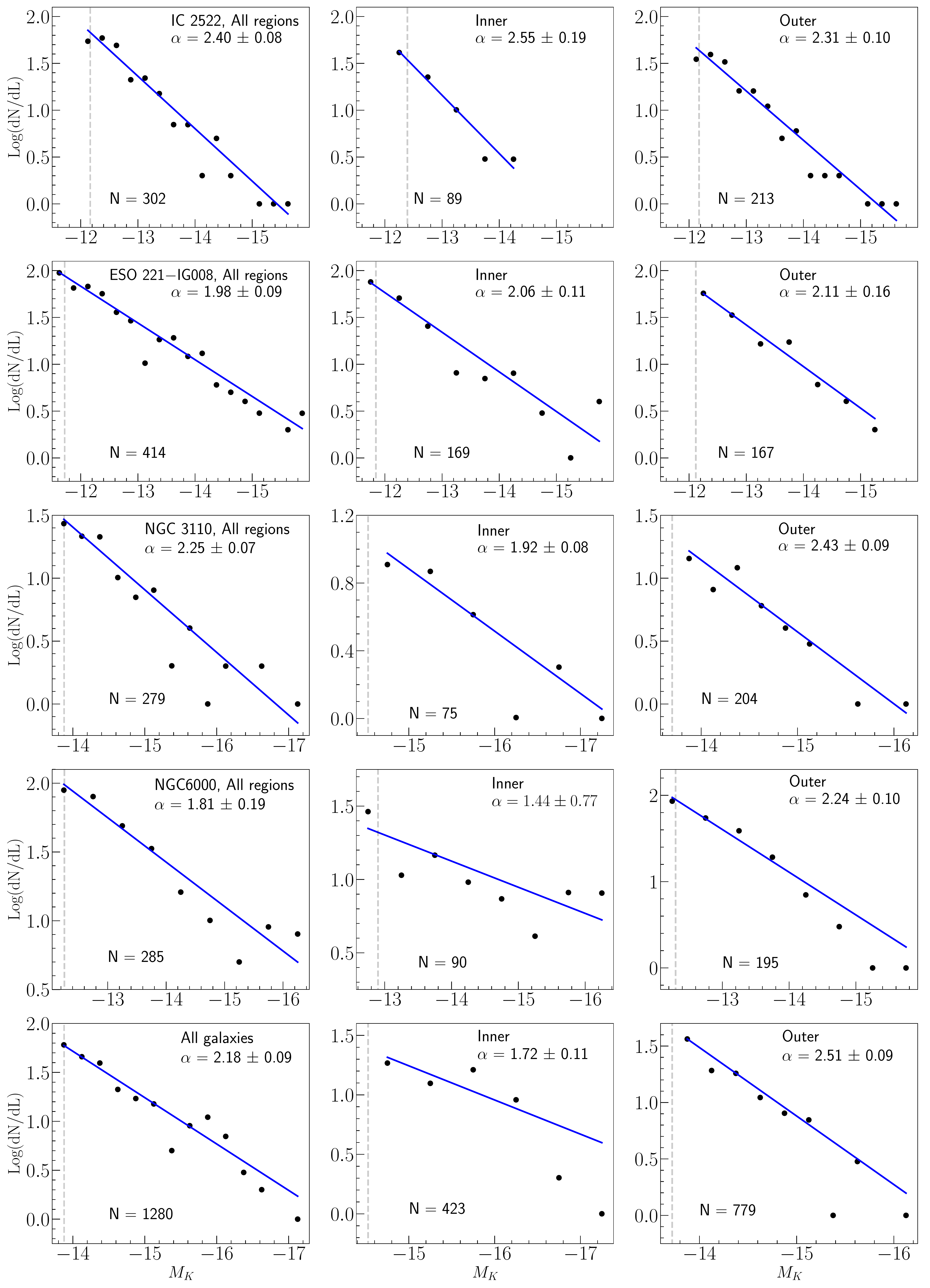}}}
\\
\caption{\small Completeness-corrected CLFs of four cluster-rich galaxies for the entire YMC population {\it (left)} and for two distinct subgroups segregated by the cluster physical locations in the field of the host galaxy: the nuclear regions {\it (middle)} and the arms or other starburst regions {\it (right)}. We also used these regions to estimate new sets of completeness fractions that are applied to the LFs. There are clearly some differences in the CLF shapes and power-law slopes for different regions of NGC 3110 and NGC 6000. The bottom panels show the corresponding composite CLFs of the cluster-rich galaxies and by putting together the catalogues associated with their inner and outer regions.  Labels are the same as in Fig. \ref{fig:CLFs}.}
\label{fig:CLF-rich}
\end{figure*}

Data points from the following targets were specifically chosen to test one more time the robustness of the completeness fractions and to ultimately derive CLFs on sub-galactic scales: IC\,2522, ESO\,221\m IG008, NGC\,3110, and NGC\,6000. We selected these galaxies that are cluster-rich, i.e. they each host $\approx$\,300 YMCs in the $K$-band, to have enough statistics and hence a robust analysis. Their luminosity distances span between 32 and 75 Mpc. In addition, this  subsample represents well the overall SUNBIRD sample in terms of morphological types: while IC\,2522 and NGC\,3110 present distinct spiral arms, NGC\,6000 is a disk galaxy with a complex nuclear region. ESO\,221\m IG008, on the other hand, has a peculiar
morphology with an irregular nuclear region (see Fig.\,\ref{fig:sample}). Since the influence of the galactic environment on the YMC properties is best studied when the star cluster masses and ages are known, we thus use the CLFs in this work as a tool to define a first order approximation of how YMCs from distinct physical regions may differ from one another. Can a sub-population of the YMCs be responsible for a power-law slope being steeper/shallower than the canonical value? Do the star clusters in the nuclear regions have similar LF trends as those from the outer regions of the galaxy? 

We divided the star cluster candidates of each target into two distinct populations depending on their projected physical locations in the galaxy: nuclear regions vs.\,spiral arms or any outer starburst regions of the galaxy. New sets of completeness fractions were then derived to construct CLFs corrected for observational incompleteness. We then re-ran the MC completeness simulations under the same conditions as in Section \ref{sec:comp}, except that the newly-defined regions were based on the visual NIR morphology of the galaxy instead of the usual equally-spaced background levels. Table \ref{tab:slope-frac} lists the 80\,percent completeness limits of these { inner and outer sub-regions (Cols 6 \& 8)}.  Fig. \ref{fig:CLF-rich} shows the derived completeness-corrected LFs for the entire galaxy (left) and for the two distinct physical regions (middle and right panels).  Although the bright end slope of the two sub-galactic LFs are slightly different in the case of IC\,2522 ($2.55\pm 0.19~{\rm vs.}~2.31 \pm 0.10$) and ESO\,221\m IG008 ($2.06\pm 0.11~{\rm vs.}~2.11 \pm 0.16$), the values of $\alpha$ remain consistent within the error estimates of the power-law fit. However, for NGC\,3110 and NGC\,6000, CLFs of the nuclear regions have shallower slopes ($\alpha = 1.92 \pm 0.08$ and $1.44 \pm 0.17$, respectively) compared to those of the outer regions ($\alpha = 2.43 \pm 0.09$ and $2.24 \pm 0.10$). { The bottom panels of Fig. \ref{fig:CLF-rich} also show the completeness-corrected composite LFs of the combined star clusters in the inner (middle) and outer (right) regions of the cluster-rich galaxies. We perform the fit until the 80 percent completeness level of NGC\,3110 to ensure that the SC catatogues of the supergalaxy are 80 percent complete. Once again, the inner region of the supergalaxy has a shallower power-law index  ($\alpha = 1.72 \pm 0.11$) compared to the slope of the outer region ($\alpha = 2.51 \pm 0.09$). The value of the slope is $\alpha = 2.18 \pm 0.09$, if we combine both regions of the supergalaxy (bottom-left panel).}  Though uncertainties and biases such as statistical and spatial resolution effects may play a role in flattening the LFs for the nuclear regions, they cannot fully explain the significant difference of $\sim 0.5 - 0.8$ in the slopes of the two YMC sub-populations (see Section \ref{sec:test-LF}). This discrepancy will be discussed further in Section \ref{sec:discuss}. 

Finally, Table\,\ref{tab:slope-frac} also compares the values of the 80\,percent completeness limit and the LF slope for the entire cluster-rich galaxy from the two different sets of completeness fractions. { These values were estimated based on regions defined with background contours levels (Col 2) and then the galaxy morphology (Col 4), respectively.}  We find that the values of these two parameters are quite similar within their uncertainties. The use of equally-spaced background levels to define different regions of the host galaxy in Section \ref{sec:comp} is therefore a reasonable approach while deriving the completeness fractions and we expect the measured values of $\alpha$ in Table \ref{tab:alpha} to be unaffected by completeness bias. 

\begin{table*}
\begin{scriptsize}
\caption{The power-law slope of cluster-rich galaxies as a function of the completeness fractions.}
\label{tab:slope-frac}
\begin{tabular}{lcccc|cccc}
\hline 
   \noalign{\smallskip}
\multicolumn{1}{l@{\hspace{0.05cm}}}{Galaxy name} &
\multicolumn{2}{c@{\hspace{0.05cm}}}{Contour levels} &
\multicolumn{2}{c@{\hspace{0.05cm}}}{Galaxy morphology} &
\multicolumn{2}{c@{\hspace{0.05cm}}}{Nuclear regions} &
\multicolumn{2}{c@{\hspace{0.05cm}}}{Outer regions} \\
   
    	& Comp.lim & $\alpha_{1} $  &   Comp.lim & $\alpha_{1}'$ & Comp.lim & $\alpha$ & Comp.lim & $\alpha$  \\
(1) & (2) & (3) & (4) & (5) & (6) & (7) & (8) & (9) \\

   \noalign{\smallskip}
\hline \hline
   \noalign{\smallskip}
IC\,2522                  & \m12.2  &  2.41 $\pm$ 0.08   &  \m12.2  & 2.40 $\pm$ 0.08 & \m12.4 & 2.55 $\pm$ 0.19 & \m12.2 & 2.31 $\pm$ 0.10 \\ 
ESO\,221\m IG008	   & \m11.9  & 2.00 $\pm$ 0.11   &  \m11.7  & 1.98  $\pm$ 0.09 & \m11.8 & 2.06 $\pm$ 0.11 & \m12.1 & 2.11 $\pm$ 0.16    \\
NGC\,3110               & \m13.8  & 2.31 $\pm$ 0.08   &  \m13.9  & 2.25 $\pm$ 0.07 & \m14.5 & 1.92 $\pm$ 0.08 & \m13.7 & 2.43 $\pm$ 0.09 \\
NGC\,6000               & \m12.5  & 1.82 $\pm$ 0.19   &  \m12.3  & 1.81 $\pm$ 0.19 & \m12.9 & 1.44 $\pm$ 0.07 & \m12.3 & 2.24 $\pm$ 0.10 \\
\noalign{\smallskip}
\hline
\noalign{\smallskip}
\multicolumn{9}{@{} p{13cm} @{}}{\footnotesize{\textbf{Notes. }Col 1: galaxy name; Cols $2 - 5$: the values of the 80 percent completeness limit and the power-law slope using two different completeness fractions that were estimated based on regions defined with background contours levels and the galaxy morphology, respectively; Cols $6 - 9$: the values of the 80 percent completeness limit and $\alpha$ considering nuclear and outer regions of the cluster-rich galaxies, respectively.}}

\end{tabular}
\end{scriptsize}
\end{table*}

\section{Other Correlation searches with the CLF slopes of individual galaxies }\label{sec:search}

In this section, we check whether there are trends between the power-law slope $\alpha$ of the cluster luminosity function and the brightest star clusters as well as the global properties of the host galaxy. We also compare our results to previous trends in the literature. We remind the reader that we use the value of $\alpha$ from a constant magnitude binning throughout the analysis. We refer to Sections \ref{sec:Ramphul-work} and \ref{sec:CLF-SFRDens} on how the galaxy global properties were derived.

\subsection{Trends with the brightest star clusters}\label{sec:alpha-brightest}

\begin{figure}
\centering
 \begin{tabular}{c}
\resizebox{1.\hsize}{!}{\rotatebox{0}{\includegraphics{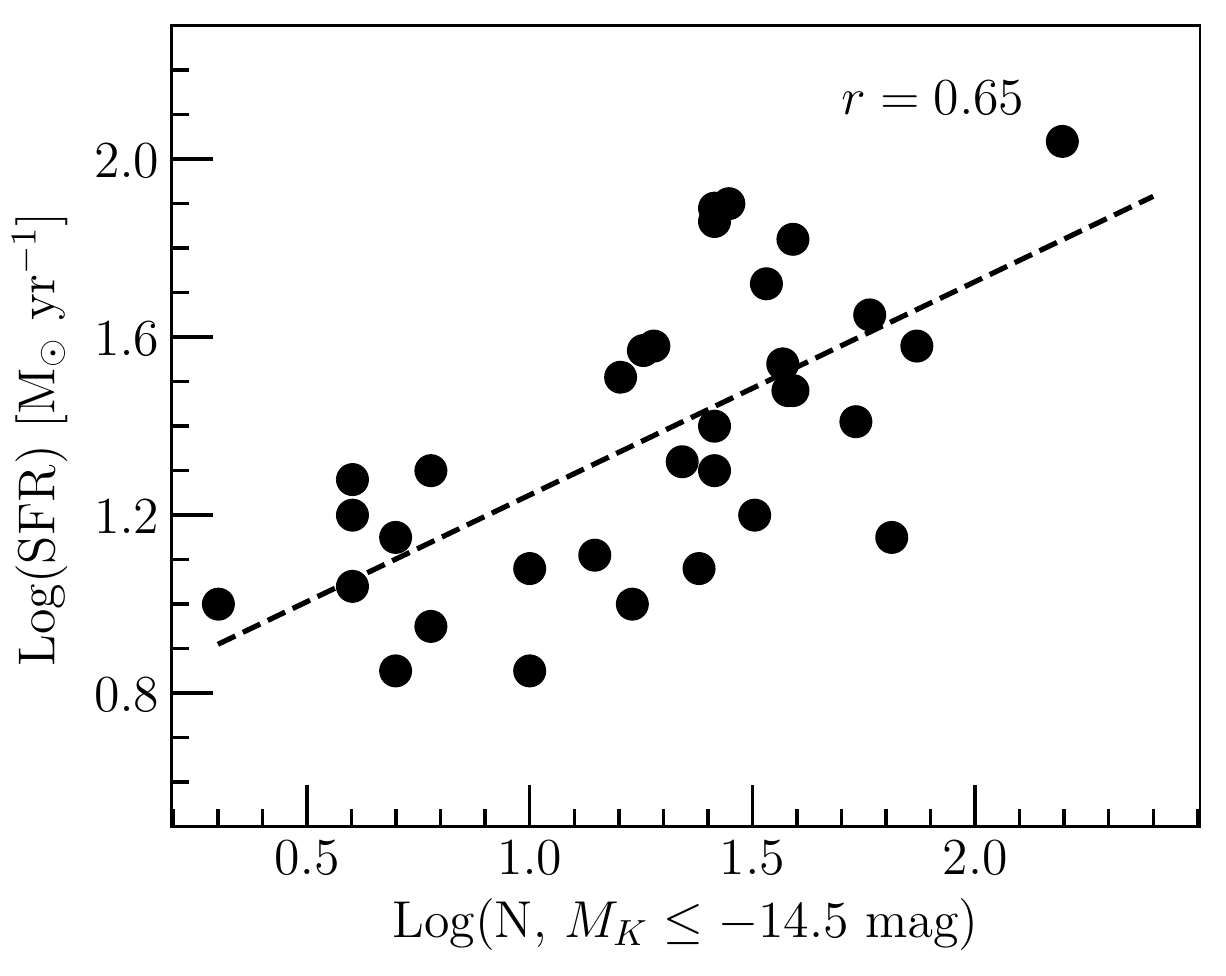}}}
\end{tabular}
\caption{\small The host galaxy SFR plotted against the number of clusters with $M_K \leq -14.5$\,mag for the SUNBIRD targets. The value of the correlation coefficient is { $r = 0.65$}.}
\label{fig:SFR-logN}
\end{figure}

The brightest star clusters are useful tools to help reconstruct the current SFH of the host galaxy { since YMCs form whenever there is intense SF activity \citep[e.g.][]{2010ARA&A..48..431P}. Because of a broadly consistent size-of-sample effect, a host environment with a higher SFR level will produce more YMCs (Fig. \ref{fig:SFR-logN}) and hence it will increase the chance of getting a brighter cluster  \citep[e.g.][]{2014AJ....147...78W}. Nevertheless, these objects are also deemed useful for checking whether physical constraints might partially} define the properties of the overall cluster population besides statistical bias. In fact, \citet{2013ApJ...775L..38R} could only explain the tightness of the brightest cluster NIR magnitude -- SFR relation of the SUNBIRD sample by considering both internal and external factors to govern the cluster formation mechanisms. MC simulations were conducted by the authors to assess the effects of pure random sampling on the magnitude of the brightest cluster and the slope of the CLF. They found that pure random sampling alone could not be the reason of the small scatter in the relation. More details of the analysis can be found in \citet{2013ApJ...775L..38R}. If the brightest cluster magnitude and SFR are tightly correlated, then one might as well expect an imprint of the host environment in the $\alpha$ vs. SFR (Section \ref{sec:alpha-SFR}) and $\alpha$ vs. brightest cluster plots.

\begin{figure*}
\centering
 \begin{tabular}{c}
\includegraphics[width=0.97\linewidth]{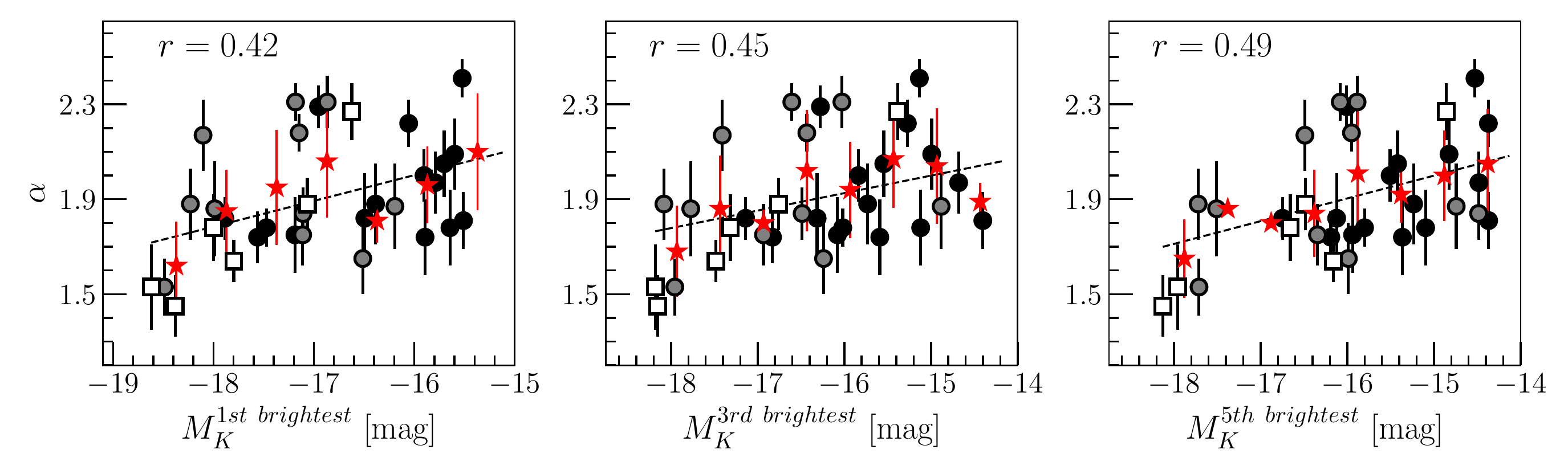}
\end{tabular}
\caption{\small The power-law slope $\alpha$ plotted against the magnitude of the first {\it (left)}, third {\it (middle)} and fifth {\it (right)} brightest star cluster candidates. Data points are labelled with respect to the distance of the host galaxy: open squares for the most distant targets where $D_L > 100$\,Mpc and circles for the $D_L \leq 100$\,Mpc targets where the ones at $D_L \leq 60$\,Mpc are black and those at $60 < D_L \leq 100$\,Mpc are grey. A constant binning of $\alpha$ results in values shown as red stars with the standard deviation of each binned data used for error bars. The dashed line represents a linear fit to these data points.}
\label{fig:alpha-brightest}
\end{figure*}

The left and right panels of Fig.\,\ref{fig:alpha-brightest} show respectively the LF slope plotted against the magnitude of the first and the fifth brightest star cluster candidates of each target. These sources are specifically chosen as they best represent the overall characteristics of the cluster population \citep{2009A&A...494..539L}. We also include the third brightest cluster (middle panel) to better study the trend of the slope - brightest star cluster relation. Since the magnitude of the first brightest cluster is likely more susceptible to stochastic effects, considering the other two brightest clusters will help derive robust analysis less affected by such bias. We notice a weak-to-mild anti-correlation between the value of $\alpha$ and the magnitude of the brightest star cluster in all three plots. The associated value of $r$ varies between 0.42 and 0.49 depending on the brightest star cluster used and the plot which considers the fifth brightest has the least dispersed data points. 

While 82 percent of the SUNBIRD first brightest clusters lie within 2 kpc from the galaxy center, there are respectively about 70 and 42 percent of the third and the fifth brightest YMCs to reside within that same region. We also plotted $\alpha$ against the nuclear distance of the brightest YMCs but did not find any obvious trend between the two values. 
 
 \citet{2014AJ....147...78W} have found a similar trend between $\alpha $ and the brightest cluster in their sample of normal spiral galaxies: the most luminous YMCs tend to be associated with shallower CLF slopes. They mainly interpret such a behaviour as a mere reflection of the existing correlations between log N, log SFR and $M^{\rm brightest}$ due to the size-of-sample effect. For the SUNBIRD sample, 12 of the first brightest clusters with magnitude $M_K < -17$\,mag, i.e. \,35\,percent, are part of a population with a CLF slope smaller than the median value $\alpha \sim 1.86$.  Based on our findings in \citet{2013ApJ...775L..38R},  we suggest possible external factors (on top of statistical bias) to partially explain such a trend (Section \ref{sec:discuss}).

\subsection{Trends with the host galaxy properties}

\subsubsection{The galaxy SFR, sSFR, $\Sigma_{\rm SFR}$ and EW(H$\alpha$) level}\label{sec:alpha-SFR}
The top panels of Fig. \ref{fig:alpha-prop} plot the LF slope against the logarithmic value of the SFR (left panel), the specific SFR (middle) and { the SFR density (right).  It is important to include all three plots while conducting correlation searches with $\alpha$ to properly disentangle the role of any physical underlying quantities from the broadly consistent size-of-sample effect. The left panel of the second row shows $\alpha$ plotted against} the EW(H$\alpha$) of the host galaxy. The computed values of $r$ are \m0.34, \m0.54, \m0.64 and 0.20, respectively. The value of $r$ becomes \m0.59 and then \m0.55 for $\alpha$ vs. $\Sigma_{\rm SFR}$ plot if we respectively consider regions defined at a level of 3$\sigma$ and 5$\sigma$ above the sky background while estimating the area where the global SFR was measured. The correlation coefficient associated with the fourth plot, i.e. $\alpha$ vs. EW(H$\alpha$), becomes $r = 0.49$ if we exclude prominent outliers with log(EW) < 0.5.

The plot of $\alpha$ vs. $\Sigma_{\rm SFR}$  exhibits the strongest correlation followed by $\alpha$ vs. sSFR and then $\alpha$ vs. SFR where galaxies with higher levels of $\Sigma_{\rm SFR}$, sSFR and SFR generally have lower values of $\alpha$. If we bin the slopes, we get the points labelled as red stars in the top panels of Fig. \ref{fig:alpha-prop}. The binned data points give respectively correlation coefficients $r$ = \m0.97, \m0.86 and \m0.97, i.e. the trend becomes even more prominent. The corresponding $p$-values are 0.035, 0.006 and 0.007, respectively. These correlations are in agreement with the derived CLF slopes from supergalaxies split as a function of SFR, sSFR and $\Sigma_{\rm SFR}$ presented in Sections \ref{sec:CLF-SFR}, \ref{sec:CLF-sSFR} and \ref{sec:CLF-SFRDens}, respectively.  A weak-to-moderate correlation between $\alpha$ and the SFR has also been reported by \citet{2014AJ....147...78W} and \citet{2016MNRAS.462.3766C}. While the former authors associate both size-of-sample effect and minor external effects with such a trend, the latter suggest that the correlation may be linked to the SF mechanisms in their sample of nearby galaxies. In fact, \citet{2016MNRAS.462.3766C}  also observed a correlation between $\alpha$ and $\Sigma_{\rm SFR}$, though they did not find any correlation in the case of $\alpha$ vs. sSFR. 

As already indicated by the value of $r$, there is a weak-to-moderate positive correlation between $\alpha$ and EW(H$\alpha$) after excluding prominent outliers. This is interesting given that we find a moderately strong anti-correlation with the galaxy global SFR. We suggest possible physical scenarios responsible for such a trend in Section \ref{sec:discuss}.

\subsubsection{Other global parameters}\label{sec:alpha-other}

\begin{figure*}
    \centering
    \begin{tabular}{c}
    \includegraphics[width=0.95\textwidth]{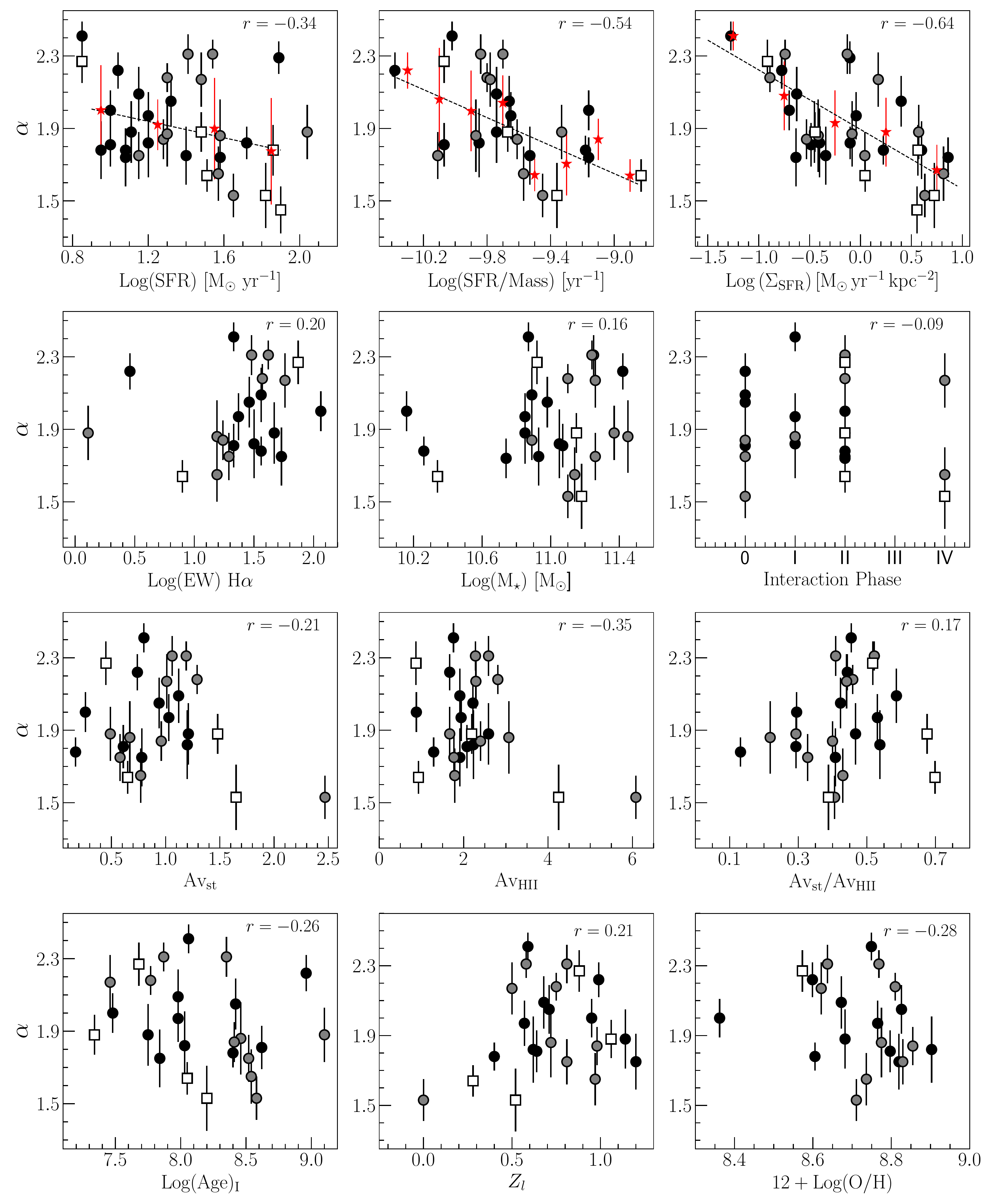}
    \end{tabular}\caption{\small Correlation search between the LF slope and the properties of the SUNBIRD sample. The significance of each correlation coefficient $r$ is given in each panel. Labels the same as in Fig.\ref{fig:alpha-brightest}. { {\em Upper panels:} $\alpha$ plotted against the host galaxy's SFR, sSFR, and $\Sigma_{\rm SFR}$. A constant binning of $\alpha$ results in values shown as red stars and the dashed line represents a linear fit to these data points. {\em Second row:} $\alpha$ vs. EW(H$\alpha$) as measured from the spectra, the stellar mass and the interaction stage of the galaxy.  {\em Third row:} $\alpha$ vs. the stellar and ionized visual extinction, and the ratio between the two quantities. {\em Lower:}  $\alpha$ vs. the light-weighted age and metallicity, and the Oxygen abundances.}}
    \label{fig:alpha-prop} 
\end{figure*}

Correlation searches between the LF slope and other physical properties of the SUNBIRD galaxies were also performed in this work. { The middle and right panels in the second row of Fig. \ref{fig:alpha-prop}} present $\alpha$ plotted against the stellar mass and the merger stage, respectively. The third row panels show plots of $\alpha$ versus the stellar and ionized visual extinction as well as the ratio between the two parameters. Finally, the bottom panels display $\alpha$ vs. the light-weighted age and metallicity as well as the Oxygen abundances 12\p Log(O/H) of the host galaxy. Overall, we find a weak to no correlation between $\alpha$ and these global parameters. The absolute value of the correlation coefficient ranges between 0.09 and 0.35.

Although the stellar mass M$_{\star}$ and the optical magnitude $M_B$ are well-known to be tightly correlated  \citep[e.g.][]{2003ApJS..149..289B}, 
there is no prominent trend between $\alpha$ and any of these two quantities.
This could be partly due to the narrow ranges of the SUNBIRD high stellar masses ($10 < {\rm log\,M}_{\star}/{\rm M}_{\odot} < 11.5$),
whereas the \citet{2016MNRAS.462.3766C} data points, for instance, are widely spread between $6 < {\rm log\,M}_{\star}/{\rm M}_{\odot} < 11$. 
Their high-mass galaxies appear to have flatter CLFs and the low luminosity ones tend to be associated with steeper CLF slopes. 

While the stellar visual extinction did not show any prominent trend with the slope $\alpha$ ($r =-0.21$), the plot $\alpha~{\rm vs.}~A_{\rm v,\,HII}$ rather hints a weak trend between these two parameters ($r = -0.35, p = 0.08)$, i.e. a steeper slope for a host galaxy with HII regions with smaller extinction. No previous works have searched for any correlation between $\alpha$ and these quantities. Despite the weak trend which is mainly driven by two highly reddened galaxies ($r$ = 0.12 excluding these outliers), we provide possible physical explanations in Section \ref{sec:discuss}. 

There is no obvious correlation between $\alpha$ and the light-weighted age ($r =-0.26$) and the light-weighted metallicity ($r = 0.21$) as well as 12\p Log(O/H) ($r =-0.28$). \citet{2016MNRAS.462.3766C} did also find no trend with the latter quantity. We also find no trend by considering the mass-weighted parameters listed in Table\,\ref{tab:prop-Ramphul}. 
Finally, the plot of $\alpha$ vs. the interaction stage   presents the poorest value in the correlation coefficient ($r = -0.09$). 

\section{Discussion}\label{sec:discuss}
\subsection{Caveats and limitations}
Besides the potential of the host environment to influence the formation mechanisms of its cluster population, this external factor is also believed to affect cluster disruption processes \citep[e.g.][]{2012MNRAS.426.3008K,2014ApJ...797L..16V,2017ApJ...843...91L, 2019MNRAS.482.2530R}. In particular, an environmentally-dependent disruption effect can manifest by flattening the CLF slope of high SFR galaxies, as low-mass clusters get more heavily disrupted than the high-mass ones. Based on the current work which analyses observations from a single filter, it is however impossible to disentangle formation effects from disruption effects. Multi-band data are needed for cluster age and mass modelling so that robust analyses like in the case of the LIRG Arp 299 \citep{2019MNRAS.482.2530R} could be conducted to define the level and the nature of the cluster disruption mechanism at play. We hope to study this aspect further in future works.

It is therefore clear that without an estimate of the cluster mass and age, we are not able to fully assess the influence of the environment on the YMC properties nor check whether older star clusters exhibit steeper or shallower LF slopes \citep[e.g.][]{1998AJ....115.1543O,2019MNRAS.482.2530R}.  Nevertheless, we herewith provide first order approximations of the ranges of these two parameters by taking into account the computed values for Arp 299 (which is a SUNBIRD target) and considering an SSP model to reconstruct the evolution of a NIR cluster brightness with time. \citet{2019MNRAS.482.2530R} found that more than 60 percent of the YMCs in Arp 299 have ages younger than 10 $-$ 15 Myr old. This is not surprising because of the extreme environments of interacting LIRGs and starburst galaxies that favor the extensive birth of YMCs \citep[e.g.][]{1999AJ....118.1551W,2020MNRAS.499.3267A}.  Such young YMCs should also be easily detected with $K$-band observations since the luminosities of high-mass stars entering the red supergiant phase are predominant at that time, and which lead to a more luminous cluster exposed in the NIR regime  (e.g. \citealp{2007ApJ...671..781D}, \citetalias{2013MNRAS.431..554R}). We thus refer to this age interval while estimating a mass range of the SUNBIRD cluster populations. If this is the case, then one would expect a weaker caveat of flattening the CLF slope being caused by disruption, since such clusters would not be old enough to be strongly affected by mass-dependent disruption, which is usually considered to act on $>$\,100 Myr time-scales \citep[e.g.][]{2006MNRAS.371..793G}. In fact, assuming the mass-to-light ratio of a $\sim$\,10 Myr old cluster, mass range approximations in  \citetalias{2013MNRAS.431..554R} revealed that NIR-selected YMCs with  $-18 < M_K\,({\rm mag}) < -13$ correspond to cluster masses between $\approx 2 \times 10^4\,{\rm M}_{\odot}$ and $\approx 4 \times 10^6\,{\rm M}_{\odot}$. This range was derived based on a {\tt Starburst99} \citep{1999ApJS..123....3L} SSP model assuming an instantaneous SF with a Kroupa IMF. Under the same assumption, the cutoff limit of $-$14.5 mag widely used in this work (where we expect most of the data to be more than 50 percent complete) is associated with a lower mass limit of $\approx  10^5\,{\rm M}_{\odot}$. This is in agreement with the completeness mass limit of $10^{4.8}\,{\rm M}_{\odot}$  from the cluster age-mass diagram of Arp 299 within a time interval of $10 - 200$ Myr \citep{2019MNRAS.482.2530R}.
We remind the reader that these age and mass ranges are simply rough estimates and a proper modelling needs to be conducted before further discussion.

In spite of these caveats and limitations, findings from this work are still relevant to help us provide a first order explanation of the flattening in the CLF slopes. Are we looking at the reflection of a size-of-sample effect or is there any other external factors that need to be considered as well?

\subsection{Systematic biases alone cannot explain the flatter slopes}

Fig. \ref{fig:CLFs} indicated that the completeness corrected CLFs of the 26 SUNBIRD targets mainly studied in this work are generally well-fitted by a single power-law function. The median and average of the derived slopes 
are equal to $\alpha_{\rm med}^{\rm cte} =1.87 \pm 0.23$ and $\alpha_{\rm aver}^{\rm cte} = 1.93 \pm 0.23$, respectively. The average slope becomes
$\alpha_{\rm aver}^{\rm cte} = 1.92 \pm 0.24 $ when taking into account the whole SUNBIRD sample, i.e. including the other 8 LIRGs from our \citetalias{2013MNRAS.431..554R} pilot study. 

To avoid any bias from low number statistics in some cases, we also constructed composite CLFs of the whole SUNBIRD sample and for targets with N $>$ 30, excluding the distant Bird galaxy, with cluster magnitudes $M_K \leq -14.5$\,mag (see Fig.\,\ref{fig:CLF-all}). The luminosity-limited cluster catalogues of { these supergalaxies have} LF slopes of $\alpha = 1.98 \pm 0.10$ { and $\alpha = 2.02 \pm 0.09$, respectively. These values} as well as the derived average slope are similar to the canonical slope $\alpha = 2$ and they are consistent with the results in our pilot study ($\alpha_{\rm aver} \approx 1.9$, \citetalias{2013MNRAS.431..554R}) and other works reporting CLFs from larger samples of LIRGs ($\alpha \approx 1.8 - 2.0$, \citealt{2011PhDT.........8V} and \citealt{2011AJ....142...79M}).  
\citet{2020ApJ...888...92L} recently constructed a composite CLF of their 48 GOALS targets (where $40 \lesssim D_L \lesssim 76$ Mpc). They derived a power-law slope of $\alpha \approx 1.5$ and associated the shallower slopes with high SFR levels (in the extreme environments of strongly-star forming galaxies) which are responsible for the overabundance of  star-forming regions hosting luminous YMCs populating the bright end of the LF. 
Nonetheless, they also warned that the flattening in the slope could be partly due to resolution effects, especially for galaxies at distances $D_L > 200$\,Mpc. Finally, \citet{2016ApJ...826...32M} have derived a CLF with $\alpha \sim 2.2$ for NGC 3256, a merging pair of galaxies with a total SFR of 50\,${\rm M_{\odot}\,yr^{-1}}$ 
at a distance of roughly 36 Mpc. Their value is similar to that derived in galaxies of lower SFR levels, and noted that a flatter measured CLF/CMF slope can also arise from various factors such as incompleteness, resolution bias, binning or inadequate selection process of the star cluster population.

As a sanity check on whether biases and uncertainties significantly impact the SUNBIRD CLFs and their power-law slopes, we therefore tested the robustness of our results in Section\,\ref{sec:test-LF}. {\em i)} We found that regardless of the binning method used (constant vs. variable), the values of $\alpha$ generally  remain shallower with its median slope ranging between $1.86 - 1.88$ for both cases. {\em ii)} Fig. \ref{fig:alpha-logN} shows that there is no clear trend, but rather a random scatter between $\alpha$ and number of YMCs { (at $M_K \leq$ 80\,\% completeness limit)} in a galaxy. This is indicative of our CLF analyses not being significantly impacted by size-of-sample effect, nor stochastic sampling for cluster-poor galaxies. {\em iii)} Regarding the reliability of the completeness rates applied to the catalogue, we derived three different sets of completeness fractions by varying the PSF size of the artificial source model (see Fig. \ref{fig:lf-psf}). The outputs from the MC simulations are consistent with the varying PSF widths and the value of $\alpha$ spans between 1.87 and 1.92 for ESO 264\m G036, one of the SUNBIRD targets used as a test bed for this sanity check. Furthermore, other versions of completeness corrected  (composite) CLFs for our cluster-rich targets (left panels of Fig. \ref{fig:CLF-rich}) confirmed that the defined regions based on the galaxy's varying background level, and its NIR morphology, return completeness limits that are consistent with each other (Table \ref{tab:slope-frac}). 
{\em iv)} Finally, we note that we had already assessed the effect of spatial resolution on the derived CLFs in the extensive blending analysis performed in \citetalias{2013MNRAS.431..554R}. The scatter plot of $\alpha$ vs. luminosity distance $D_L$ where $r = -0.17$ and the value of $\alpha \approx 2$ of the composite CLF for targets with distances $D_L \leq$ 100 Mpc in Fig.\,\ref{fig:alpha-DL} reinforce our conclusion that blending effects have no significant impact on the derived CLFs, in particular at distances below 100 Mpc for the SUNBIRD sample.

Based on these various investigations, we conclude that size-of-sample effect alone and/or other aforementioned biases are not mainly responsible for the observed flatter LF slopes. We thus explore the possibility of external factors influencing the SC properties of the SUNBIRD sample.

\subsection{Host environments appear to influence the CLF slopes}

CLFs of three subsamples split as a function of the galaxy's global SFR clearly indicate that the smaller values of $\alpha$ are associated with galaxies of higher SFR levels (see Fig. \ref{fig:LFs-SFR}). The value of $\alpha$ clearly gets shallower by $\approx$\,0.5 as we compare CLF indices between subsamples with ${\rm SFR} \leq 30\,{\rm M_{\odot}\,yr^{-1}}$ ($\alpha = 2.27 \pm 0.08$) and ${\rm SFR } > 60\,{\rm M_{\odot}\,yr^{-1}}$ ($\alpha = 1.74 \pm 0.09$). Such a change in the value of $\alpha$ is also observed in Fig. \ref{fig:LFs-sSFR} when drawing CLF composites from subsamples with log\,(SFR/M$_{\star}) \leq -9.68\,{\rm yr^{-1}}$ ($\alpha = 2.22 \pm 0.08$) and log\,(SFR/M$_{\star}) > -9.68\,{\rm yr^{-1}}$ ($\alpha = 1.82 \pm 0.09$), i.e. $\Delta \alpha \approx 0.4$. Galaxies with higher specific SFRs are associated with smaller values of $\alpha$ as well. { Finally, we also find a similar trend in Fig. \ref{fig:LFs-SFRDens} when fitting a power-law slope to the CLF composites of subsamples with log\,($\Sigma_{\rm SFR}) \leq -0.10\,{\rm M_{\odot}\,yr^{-1}\,kpc^{-2}}$ ($\alpha = 2.20 \pm 0.12$) and log\,($\Sigma_{\rm SFR}) > -0.10\,{\rm M_{\odot}\,yr^{-1}\,kpc^{-2}}$ ($\alpha = 1.82 \pm 0.10$), i.e. $\Delta \alpha \approx 0.4$. The CLF flattens with an increasing SFR density.} 

Analyses on sub-galactic scales of cluster-rich galaxies  (either considering individual galaxies or composite supergalaxies) also show that YMCs in the nuclear regions generally have shallower LF slopes than the cluster sub-populations hosted by the other parts of the galaxy (see middle and right panels of Fig. \ref{fig:CLF-rich}). For NGC\,3110 and NGC\,6000, we record a difference of $\alpha > 0.5$ in the CLF slopes of these two distinct regions. External factors such as environmental effects could partly explain this large discrepancy, especially given that SF activity is usually more intense in the galactic nuclear regions \citep[e.g.][]{2012MNRAS.419.2606B,2014MNRAS.440L.116S, 2017ApJ...841..131A,2019MNRAS.482.2530R}. To better quantify the role of SF activity on sub-galactic scales,  spectroscopic observations can be used to spatially disentangle the SFR contribution from each region based on the galaxy NIR image but such analysis is, however, beyond the scope of this work.  Nevertheless, ALMA and Submillimeter Array observational studies of molecular gas (traced by CO emission) in NGC 3110 \citep{2018ApJ...866...77E, 2022arXiv220310303K} and NGC 6000  \citep{2010AJ....139.2241M} revealed that there is a high gas concentration in the nuclear regions of these two intensely star-forming galaxies (more than 85 percent of the emission for NGC 6000). Such distributions are in agreement with these regions hosting a large number of YMCs. 

The findings in this work hence seem to suggest that the host galaxy environment (via its SFR { and $\Sigma_{\rm SFR}$} levels) plays a role in defining the properties of its star cluster population. This is also supported by the weak-to-moderate correlation between $\alpha$ and  both the galaxy's SFR ($r=-0.34$ { and by binning the slopes $r=-0.97$, top left panel of Fig. \ref{fig:alpha-prop}) and its SFR density ($r=-0.64$ and by binning the slopes $r=-0.97$, top right panel)}: a shallower CLF slope for a higher SFR { (density)} level. As already mentioned in Section \ref{sec:alpha-SFR}, statistical bias and size-of-sample alone cannot explain the observed correlation. Such behaviour is most likely another evidence of the influence of some external factors on the YMC characteristics (see e.g. \citealt{2016MNRAS.462.3766C}). 
The proposed argument is { partially} justified by the existence of a relation between the host galaxy SFR and log N shown in Fig. \ref{fig:SFR-logN}. 
By studying nearby spiral galaxies, \citet{2000A&A...354..836L} have reported that a rich population of YMCs is indeed generally expected in environments with a high level of SF activity, though these peculiar objects can also be found in various types of galaxies. { However, } this ubiquity is also thought to be a consequence of some underlying process such as the mean gas density that controls both the SFR and the star cluster formation, at least to a first order approximation, and hence the strong spatial correlation between YMCs and molecular gas emission (see e.g. \citealt{2017A&A...601A.146C,2018ApJ...866...77E, 2018MNRAS.481.1016G,2022arXiv220310303K}). In fact, \citet{1959ApJ...129..243S} had already proved the existence of a tight correlation between the SFR and the gas density which is expressed as $\Sigma_{\rm SFR} \propto \Sigma_{\rm gas}^N$ (where N $\sim$ 1.4, \citealt{1998ApJ...498..541K}). Environments with high gas surface density that are generally of higher interstellar medium pressure are thus ideal nurseries of YMCs since such conditions mainly govern the birth of strongly bound clusters \citep{1997ApJ...480..235E}. In addition, while a log N - SFR relation is broadly consistent with a size-of-sample effect and thus can also be seen in the case of a constant CFE, the tightness of the correlation between the brightest cluster and the SFR \citep{2013ApJ...775L..38R} as well as the dependence of CFE on the SFR density \citep{2000A&A...354..836L} could also partially explain this observed relation. The CFE - $\Sigma_{\rm SFR}$ trend has also been reported by e.g. \citet{2010MNRAS.405..857G}, \citet{2016ApJ...827...33J} and \citet{2020MNRAS.499.3267A} and theoretically predicted by e.g. \citet{2012MNRAS.426.3008K}. Such a dependence implies that gravitationally bound clusters easily form in high pressure environments with extensive starburst activity per unit area.
For interacting galaxies such as LIRGs, tidal torques may also play a role in migrating dense gas towards the nuclear regions of these systems and thus ultimately enhancing the galaxy SF activities \citep[e.g.][]{2013A&A...553A..72K}.
All these scenarios seem to reasonably provide physical explanations of the existing correlation between $\alpha$ and  both SFR and $\Sigma_{\rm SFR}$. 
Nevertheless, the physical relation is not yet fully understood -- hence the long-lasting debate in the literature \citep[e.g.][]{2017ApJ...849..128C,2019ARA&A..57..227K,2020MNRAS.494..624K,2020MNRAS.499.3267A}.

\subsection{Further evidence of the suggested influence based on the trends with various host global properties}

We also looked for other host galaxy parameters indicative of the star formation rate that might correlate with the LF slope: we plotted $\alpha$ against sSFR and then against the EW of H$\alpha$ emission in Fig. \ref{fig:alpha-prop}. The moderate-to-strong correlation of the former plot ($r = -0.54$ for all data points and $r = -0.86$ by binning the slopes) is in agreement with the observed trend between $\alpha$ and log SFR (density).  
The latter correlation is also particularly interesting as it would appear to be in contradiction to the $\alpha$ vs. SFR  (density) correlation, since higher H$\alpha$ emission obviously indicates stronger SF. However, we interpret this correlation in conjunction with the anti-correlation, albeit weak, seen between H$\alpha$ and the light-weighted age of the host galaxy in the same Figure, i.e.\ galaxies with computed older ages show generally lower $\alpha$. It should be remembered that H$\alpha$ traces the very youngest SF of the order of less than 10 Myr, whereas the SF rates discussed in this work are derived from far-IR emission tracing somewhat older SF at the $<100$ Myr scale, or even older, depending on exact SF histories \citep[e.g.][]{2012ARAA..50..531K}. Hence, both these correlations suggest that whatever effect is responsible for the flatter CLF slopes of higher SFR (density) galaxies, it would be a process that is not {\em yet} evident, or operating, at the very youngest galaxies, or portions thereof. This could, for example, suggest that the flatter cluster LF may be related to selective cluster disruption effects.  

We performed correlation searches between $\alpha$ and other global properties of the host galaxy as well (see Fig. \ref{fig:alpha-prop}). Interestingly, no correlation is seen with the interaction stage in our sample. There is weak anti-correlation seen with extinction, though this is strongly dependent with a mere two highly reddened galaxies. Nevertheless, the apparent trend of a flatter slope for the more extinguished galaxies could be indicative of the abundance of newly-formed SCs younger than $2-5$\,Myr that are still highly embedded in the dusty ionised gas in HII regions { \citep[e.g.][]{2002AJ....124..166A,2011ApJ...729...78W,2015MNRAS.449.1106H,2017A&A...601A.146C,2019MNRAS.483.4707G,2021ApJ...909..121M}}. We suspect that such correlations would be better to study with sub-regions of galaxies rather than with their integrated properties.
 
Finally, with a correlation factor ranging between 0.42 and 0.49, we find a mild anti-correlation between the LF slope $\alpha$ and the $K$-band absolute magnitude of the first, the third, and the fifth brightest YMCs (Fig. \ref{fig:alpha-brightest}). While a size-of-sample effect is mainly responsible for the existing trend \citep{2014AJ....147...78W},
we also suggest that the $\alpha - M^{\rm brightest}$ relation can { partly} arise because of physical reasons, in particular since the most luminous YMCs are generally the youngest ones, especially for a large population of YMCs \citep{2009A&A...494..539L}. In fact, theoretical and observational studies by e.g. \citet{2008MNRAS.390..759B} and  \citet{2013ApJ...775L..38R} have agreed that the tight $M^{\rm brightest} - $ SFR relation is an imprint of the current SFR of the host galaxy and it may also occur because of the influence of the global SF properties on the luminosity of the brightest YMCs (see Section \ref{sec:alpha-brightest}). High SFR galaxies (associated with shallower $\alpha$) are thus expected to host the most luminous and youngest YMCs abundantly  (see e.g. \citealp{2020ApJ...888...92L}). 

\subsection{Imprints of the cluster formation mechanisms on the CLF slopes?}

Overall, the results and analyses in this work generally support the idea that high SFR galaxies (such as our sample) exhibit shallower CLFs slopes than gas-poor galaxies. This trend could be due to the dependence of the cluster formation (and evolution) mechanisms on the galaxy host local/global environment \citep[e.g.][]{2018MNRAS.477.1683M}: an extreme environment with strong SF activity that favors the extensive birth of YMCs, and vice versa.
By producing more luminous YMC candidates, such external effects are translated into the flattening of the CLF bright end for galaxies like those in the SUNBIRD sample. We clearly noticed this environmental dependence while analysing the YMC masses and ages of the ongoing merger Arp 299 \citep{2019MNRAS.482.2530R}.  And in addition, there may also be a contribution of a flattening of the slope at the faint end by {\em disruption} of the less massive clusters in extreme environments \citep[see e.g.][]{2014ApJ...797L..16V}. Such selective disruption is also suggested by our finding that while high SFR galaxies such as LIRGs globally are indeed associated with flat LF slopes, those galaxies that show the very youngest SF, interestingly tend to (still?) have steeper CLFs.  
Note, however, that there are other works that favor the universality of the cluster formation process, which is independent of the SF environments. While various authors have suggested mere biases and size-of-sample effect to be the main reasons behind any deviation from the canonical LF/MF slope \citep[e.g.][]{2016ApJ...826...32M,2017ApJ...849..128C,2017ApJ...843...91L,2019ApJ...872...93M,2019MNRAS.484.4897C}, this is most likely not the case for the CLFs of the SUNBIRD sample (see Section \ref{sec:test-LF}).

\section{Conclusions}\label{sec:conclusion}
This article reports a follow-up study of the $K$-band luminosity functions of YMCs in intensely star-forming galaxies. The work is aimed at improving the constraints on the star cluster formation mechanisms and our understanding of SF in general. Our findings based on the larger SUNBIRD sample of 26 targets, with ${10.6 < {\rm log}\,(L_{\rm IR}/L_{\odot}) < 11.7}$, that were imaged with the VLT/NACO AO instrument, are in agreement with pilot study results in  \citetalias{2013MNRAS.431..554R} yielding a median slope of $1.87 \pm 0.23$. There is likely an imprint of the galactic environments on the properties of YMCs hosted by galaxies with high SFR levels such as starburst galaxies and LIRGs, especially at distances $D_L < 100$\,Mpc. The observed slope of $\alpha \sim 1.9$ in most of the SUNBIRD CLFs, combined with the existing trend between the power-law slope and the galaxy SFR properties (global SFR level, specific SFR and SFR density), and the moderate $\alpha - M^{\rm brightest}_K$ relation lead us to suggest that {\it the value of $\alpha$ partially reflects the environment where the YMCs reside in}. Such a statement has been drawn after conducting an extensive assessment of all possible systematic biases and uncertainties. These are expected to not significantly affect the cluster luminosity distribution and hence the value of $\alpha$ in this work. 

Nevertheless, given the caveats of working with a single filter, and comparing our NIR-based results with CLF studies of galaxies with less intense SF activity conducted in the optical regime, future work could focus on a comprehensive investigation of the YMC properties based on multi-wavelength observations of the SUNBIRD targets. We can then compute both the age and mass of each star cluster to accurately constrain the cluster formation and disruption mechanisms. Using observations taken with the new and upcoming generation of AO imagers such as  GeMS/GSAOI on Gemini South or AOF/ERIS on VLT will also help to achieve a higher PSF resolution (down to $\sim$ 0.07 arcsec,  \citealp{2018MNRAS.473.5641K}) and hence a more efficient detection of our sources of interest.

\section*{Acknowledgements}

We thank the anonymous reviewer for his/her detailed suggestions and insightful comments to improve the clarity of this paper. Based on observations made with the ESO Telescopes at the La Silla Paranal Observatory, Chile under programmes 084.D-0261, 086.B-0901, 087.D-0444. and 089.D-0847 for main data and programmes 072.D-0433 and 073.D-0406 for earlier NACO data. Based in part on observations obtained at the Gemini Observatory, which is operated by the Association of Universities for Research in Astronomy, Inc., under a cooperative agreement with the NSF on behalf of the Gemini partnership: the National Science Foundation (United States), the National Research Council (Canada), CONICYT (Chile), Ministério da Ciência, Tecnologia e Inovação (Brazil), and the Ministerio de Ciencia, Tecnología e Innovación Productiva (Argentina). Gemini observations were taken as part of programmes GN-2008A-Q-38, GN-2008B-Q-32, GN-2009A-Q-12, GN-2009B-Q-23 and GN-2010A-Q-40 and GN-2008A-Q- 38, GN-2008B-Q-32, GN-2009A-Q-12, GN-2009B-Q-23, GN-2010A-Q-40. Some of the observations reported in this paper were obtained with the Southern African Large Telescope (SALT) under the following programs: 2011-3-RSA\_OTH-023, 2012-1-RSA\_OTH-032, 2012-2-RSA\_OTH-015, 2013-1-RSA\_OTH-024, 2013-2-RSA\_OTH-006, 2014-1-RSA\_OTH-002. 
This research was supported by the South African Radio Astronomy Observatory and the South African Astronomical Observatory, which are facilities of the National Research Foundation, an agency of the Department of Science and Innovation. ZR also acknowledges funding from the L'Or{\'e}al - UNESCO For Women In Science sub-Saharan Africa regional Programme. JK acknowledges financial support from the Academy of Finland, grant 311438.

\section*{Data availability}
All the raw VLT data underlying this article are publicly available from ESO Science Archive Facility. The final catalogues of YMC candidates will be shared on reasonable request to the corresponding author.

\bibliographystyle{apj}


\appendix
\section{}
\begin{figure*}
\renewcommand{\footnoterule}{}
\centering
\begin{tabular}{c}
\resizebox{1.\hsize}{!}{\includegraphics{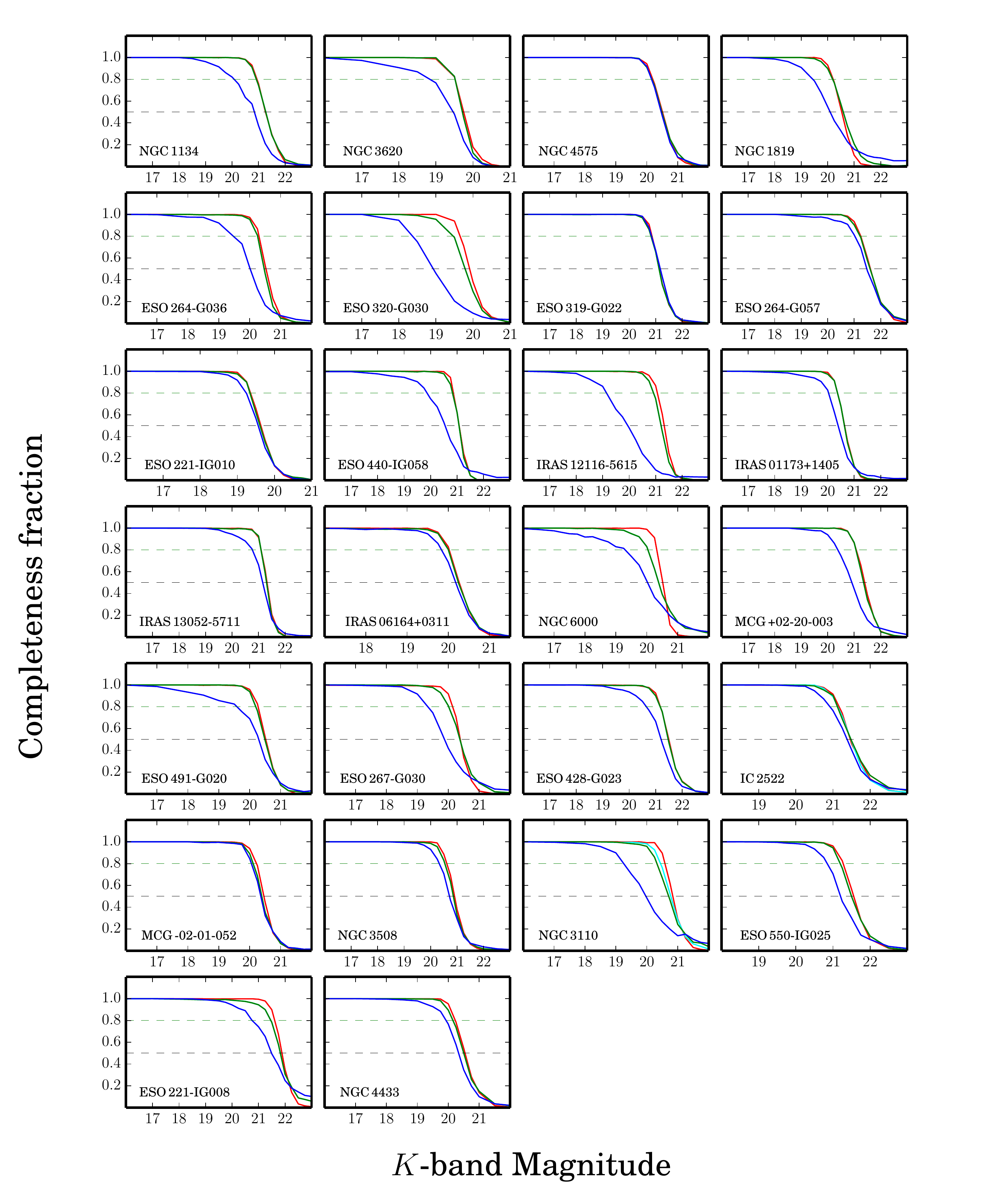}}
\end{tabular}

\caption{\small The results of Monte Carlo completeness simulations for the 26 VLT/NACO targets within regions of different background levels. The blue, green, and { red} solid lines correspond to the innermost, middle, and outer regions. IC 2522 and NGC 3110 have two middle regions to account for a more complex varying background in their field. { These two regions are represented by the green and cyan solid lines for the two targets.} The 50 and 80 percent completeness limits of the \textgravedbl middle\textacutedbl\,regions are shown as the horizontal dashed lines. }
\label{fig:comp-frac}
\end{figure*}
\label{lastpage}
\end{document}